\documentclass[iop,revtex4]{emulateapj}
\usepackage{lscape}

\shorttitle{NEP FLAMINGOS NIR Catalog}
\shortauthors{Jeon et al.}

\begin{document}

\title{$J$ and $H-$band Imaging of {\it AKARI} North Ecliptic Pole Survey Field}

\author{
Yiseul Jeon\altaffilmark{1,5}, 
Myungshin Im\altaffilmark{1,5}, 
Eugene Kang\altaffilmark{1,2,5},
Hyung Mok Lee\altaffilmark{1}, 
and 
Hideo Matsuhara\altaffilmark{3,4} }

\email{ysjeon@astro.snu.ac.kr \& mim@astro.snu.ac.kr}

\altaffiltext{1}{Astronomy Program, Department of Physics and Astronomy, Seoul National University, 1 Gwanak-ro, Gwanak-gu, Seoul 151-742, Korea}
\altaffiltext{2}{Korea Educational Broadcasting System, 2748 Nambusunhwan-ro, Gangnam-gu, Seoul 135-854, Korea}
\altaffiltext{3}{Institute of Space and Astronautical Science, Japan Aerospace Exploration Agency, 3-1-1 Yoshinodai, Chuo-ku, Sagamihara 252-5210, Japan}
\altaffiltext{4}{Department of Space and Astronautical Science, The Graduate University for Advanced Studies,  3-1-1 Yoshinodai, Chuo-ku, Sagamihara 252-5210, Japan}
\altaffiltext{5}{Visiting Astronomer, Kitt Peak National Observatory, National Optical Astronomy Observatory, which is operated by the Association of Universities for Research in Astronomy (AURA) under cooperative agreement with the National Science Foundation.}

\begin{abstract}
We present the $J$ and $H-$band source catalog covering the {\it AKARI} North Ecliptic Pole field. Filling the gap between the optical data from other follow-up observations and mid-infrared (MIR) data from {\it AKARI}, our near-infrared (NIR) data provides contiguous wavelength coverage from optical to MIR. For the $J$ and $H-$band imaging, we used the FLoridA Multi-object Imaging Near-ir Grism Observational Spectrometer (FLAMINGOS) on the Kitt Peak National Observatory 2.1m telescope covering a 5.1 deg$^2$ area down to a 5$\sigma$ depth of $\sim$21.6 mag and $\sim$21.3 mag (AB) for $J$ and $H-$band with an astrometric accuracy of 0.14$\arcsec$ and 0.17$\arcsec$ for 1$\sigma$ in R.A. and Decl. directions, respectively. We detected 208,020 sources for $J-$band and 203,832 sources for $H-$band. This NIR data is being used for studies including analysis of the physical properties of infrared sources such as stellar mass and photometric redshifts, and will be a valuable dataset for various future missions.
\end{abstract}


\section{ INTRODUCTION }

\begin{figure*}
\epsscale{1.1}
\plotone{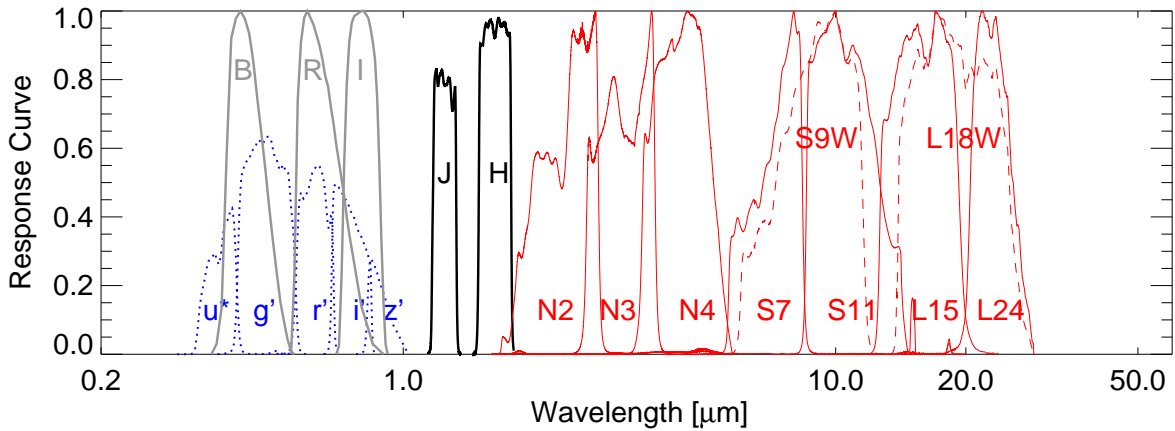}
\caption{Filter response curves of multi-wavelength data covering the NEP field. The thick dotted blue lines represent the $u^*, g', r', i'$, and $z'$ band filters from MegaCam of CFHT. The thick gray lines are for the $B$, $R$, and $I-$band filters from Maidanak observatory. The thin solid red lines and thin dashed red lines are for the {\it AKARI} IR band filter coverage ($N2, N3, N4, S7, S9W, S11, L15, L18W,$ and $L24$). The thick black lines are for our $J$ and $H-$band filters, which make the wavelength coverage between the optical data and the {\it AKARI} data continuous. Table \ref{t11} lists the information of multi-wavelength data for the filters, shown here.
\label{f01}}
\end{figure*}

{\it AKARI} is an infrared (IR) space telescope launched in 2006 \citep{mura07} which has conducted an all-sky survey in far-IR (FIR) as well as imaging observations of regions of special interest from near-IR (NIR) to FIR. Since {\it AKARI} was placed on a Sun-synchronous orbit, it conducted deep observations only toward the north and the south ecliptic poles. Observing these areas in the contiguous nine passbands between 2.4 and 26$\micron$ covering NIR to mid-IR (MIR), {\it AKARI} provided valuable NIR to MIR imaging survey data where the {\it Spitzer Space Telescope} has limitations, especially at the 11, 15, and 18$\micron$ bands. 

Taking the advantage of the Sun-synchronous orbit, {\it AKARI} carried out deep surveys with nine filters of the InfraRed Camera (IRC)  that cover from 2 to 26 $\mu$m contiguously toward the region around the North Ecliptic Pole \citep[NEP;][]{mats06,wada07,lee07, lee09}. 
The NEP survey is composed of two main programs, NEP-Wide and NEP-Deep surveys. The NEP-Wide survey \citep{lee09,kim12} covers the entire 5.4 deg$^2$ to the depths of 20.9 (AB mag) to 17.8 (AB mag) from 2.4 to 24 $\mu$m (5$\sigma$ point-source sensitivity), whereas the NEP-Deep survey \citep{wada08,taka12} covers a 0.67 deg$^2$ area with exposures 3--10 times longer than the NEP-Wide survey. With its continuous MIR wavelength range and wide field of view, {\it AKARI} NEP survey provides a unique dataset for the study of galaxy evolution, and it has been used to unveil the link between the star formation and the active galactic nucleus activity of distant galaxies \citep{karou14, hana12}, the cosmic star formation history beyond z $> 1$ \citep{goto10}, the properties of polycyclic aromatic hydrocarbons luminous galaxies \citep{taka10, ohya07}, the environmental dependence of star formation activities of galaxies in a supercluster and MIR-excess of early-type galaxies \citep{ko12}, and the nature of sources in 11 and 15 $\mu$m limited samples \citep{lee07,mats07,pear10}. Derivation of the accurate local luminosity functions in MIR is under progress (Kim et al., in preparation), by utilizing the subsequent spectroscopic survey data \citep{shim13}.

For the {\it AKARI} NEP field, a considerable amount of ground-based ancillary datasets have been obtained. Figure \ref{f01} shows the filter response curves of the multi-wavelength datasets used in the {\it AKARI} NEP field survey. Follow up imaging data observed by the MegaCam of the Canada-France-Hawaii Telescope (CFHT) with $u^*, g', r', i'$, and $z'$ filters covers 2 deg$^2$ of the central region of the NEP field \citep[][thick dotted blue lines in Figure \ref{f01}]{hwan07}. Also for the remaining area, \citet{jeon10} provided optical data of $B$, $R$, and $I-$band covering 4.9 deg$^2$ (thick gray lines in Figure \ref{f01}) using the Seoul National University 4k $\times$ 4k Camera \citep[SNUCAM;][]{im10} on the 1.5m Ritchey-Chr\'etian, AZT-22 telescope at the Maidanak observatory. Spectroscopic data of 1796 sources were obtained using Hectospec of MMT and Hydra of the WIYN telescope \citep{shim13}. The radio data were taken with the Westerbork Radio Synthesis Telescope (WSRT) over a 1.7 deg$^{2}$ area around NEP \citep{white10}, and ultraviolet (UV) imaging data were obtained with the {\it  Galaxy Evolution Explorer (GALEX)} at Far-UV and Near-UV (Malkan et al., in preparation). More recently, the entire {\it AKARI} NEP field has been covered by the Spectral and Photometric Imaging
Receiver \citep[SPIRE;][]{grif10} and a part of the NEP area is covered by the Photoconductor Array Camera and Spectrometer \citep[PACS;][]{pogl10} of the {\it Herschel Space Observatory}  \citep{serj12}. In addition, the NEP-Deep area has been observed for optical/NIR images with the Subaru telescope and the CFHT \citep{oi14} and in optical spectroscopy with the DEep Imaging Multi-Object Spectrograph (DEIMOS) on the Keck telescope (Takagi et al., in preparation).  
We summarized the information of the multi-wavelength surveys for the optical, NIR and MIR bands in Table \ref{t11} for NEP-Wide and Table \ref{t12} for NEP-Deep. 

The {\it AKARI} data \citep[][thin solid red lines and thin dashed red lines in Figure \ref{f01}]{kim12} covers a wavelength range from 2.4 to 24$\micron$ with nine filters ($N2, N3, N4, S7, S9W, S11, L15, L18W,$ and $L24$, where the number indicates the central wavelength of the filter in $\micron$), while the ground-based imaging data offering the coverage at $\lambda < 0.9$ $\micron$. 
However there is a gap in the wavelength coverage between the optical $I-$band (0.85$\micron$) and the shortest {\it AKARI} band (2.4$\micron$). Filling this gap in the spectral energy distribution (SED) will significantly enhance the value of the NEP dataset in many ways. First, by adding the NIR data, characterization of the SEDs with a contiguous wavelength range from NIR to MIR will increase the accuracy in estimating stellar masses of low redshift galaxies, because the shorter (or longer) wavelength data may be contaminated by star formation and dust content. Second, the NIR coverage can help improve the photometric redshift estimation by sampling crucial wavelengths at which some prominent features, such as the 1.6$\micron$ bump or the 4000\AA ~break, may reside for low redshift or high redshift galaxies, respectively.

The NEP field has served as a major survey field for extragalactic astronomy for {\it AKARI}, and {\it Herschel} missions, but it is also poised to do so in future missions such as {\it Euclid} \citep{laur11}, Space Infrared Telescope for Cosmology and Astrophysics \citep[{\it SPICA};][]{serj12}, Cosmic Infrared Background Experiment \citep[{\it CIBER};][]{bock13}, extended ROentgen Survey with an Imaging Telescope Array \citep[eROSITA;][]{merl12}, and LOw-Frequency ARray \citep[LOFAR;][]{vanh13}. In order to fill the gap in the wavelength coverage and to enhance the scientific value of the {\it AKARI} NEP survey, we conducted an imaging observation of the entire NEP survey area in $J$, and $H$-band (thick black lines in Figure \ref{f01}). Here, we present the $J$ and $H-$band data of the {\it AKARI} NEP field. This paper is organized as follows. In Section \ref{sec:obser}, we describe the details of the observations. In Section \ref{sec:datar}, data reduction processes are explained. In Section \ref{sec:catpr}, we describe the source detection and photometry along with the properties of the data and provide the $J$ and $H-$band merged catalog. In Section \ref{sec:datap}, we provide the result from our data such as source number counts and color-magnitude diagrams. We summarize our results in the final section. Throughout this paper, we use a cosmology with $\Omega_M=0.3, \Omega_\Lambda=0.7,$ and $H_0$ = 70 km s$^{-1}$Mpc$^{-1}$.  We use the AB magnitude system.

\begin{deluxetable*}{lcccccr}
\tablecolumns{7}
\tablewidth{0pc}
\tablecaption{Information of the multi-wavelength survey for the optical, NIR and MIR bands in NEP-Wide survey.\label{t11}}
\tablehead{
\colhead{Instrument} & \colhead{Area}   & \colhead{Pixel Size}    & \colhead{Filter} &
\colhead{Effective Wavelength}   & \colhead{Sensitivity}    & \colhead{Reference}\\
\colhead{} & \colhead{(deg$^2$)}   & \colhead{($\arcsec$ pix$^{-1}$)}    & \colhead{} &
\colhead{($\micron$)}   & \colhead{(5$\sigma$, AB mag)}    & \colhead{}
}
\startdata
MegaCam (CFHT)&2&0.19&$u*$&0.37&25.8&\citet{hwan07}\\
                           &  &       &$g'$ &0.49&25.9& \\
&&&$r'$&0.63&25.4&\\
&&&$i'$&0.77&24.6&\\
&&&$z'$&0.89&23.7&\\
\hline
SNUCAM (Maidanak 1.5m)&4.9&0.27&$B$&0.44&23.4&\citet{jeon10}\\
&&&$R$&0.64&23.1&\\
&&&$I$&0.79&22.3&\\
\hline
FLAMINGOS (KPNO 4m)&5.1&0.30&$J$&1.25&21.6&This work\\
&&&$H$&1.64&21.3&\\
\hline
IRC (AKARI)&5.4&1.46$\times$1.46&$N2$&2.4&20.9&\citet{kim12}\\
&&&$N3$&3.2&21.1&\\
&&&$N4$&4.1&21.1&\\
\cline{3-6}
&&2.34$\times$2.34&$S7$&7.0&19.5&\\
&&&$S9W$&9.0&19.3&\\
&&&$S11$&11.0&19.0&\\
\cline{3-6}
&&2.51$\times$2.39&$L15$&15.0&18.6&\\
&&&$L18$W&18.0&18.7&\\
&&&$L24$&24.0&17.8&

\enddata
\end{deluxetable*}

\begin{deluxetable*}{lcccccr}
\tablecolumns{7}
\tablewidth{0pc}
\tablecaption{Same as Table \ref{t11} but in NEP-Deep survey.\label{t12}}
\tablehead{
\colhead{Instrument} & \colhead{Area}   & \colhead{Pixel Size}    & \colhead{Filter} &
\colhead{Effective Wavelength}   & \colhead{Sensitivity}    & \colhead{Reference}\\
\colhead{} & \colhead{(deg$^2$)}   & \colhead{($\arcsec$ pix$^{-1}$)}    & \colhead{} &
\colhead{($\micron$)}   & \colhead{(5$\sigma$, AB mag)}    & \colhead{}
}
\startdata
MegaCam (CFHT)&1&0.19&$u*$&0.37&24.6&\citet{taka12}\\
\cline{4-7}
&&&$g'$&0.49&26.5&\citet{oi14}\\
&&&$r'$&0.63&25.7&\\
&&&$i'$&0.77&24.9&\\
&&&$z'$&0.89&23.9&\\
\hline
SuprimeCam (Subaru)&0.255&0.2&$B$&0.43&28.4&\citet{taka12}\\
&&&$V$&0.54&28.0&\\
&&&$R$&0.65&27.4&\\
&&&$i'$&0.80&27.0&\\
&&&$z'$&0.91&26.2&\\
\hline
WIRCam (CFHT)&0.6&0.3&$Y$ &1.02&23.2&\citet{oi14}\\
&&&$J$&1.25&22.8&\\
&&&$K_s$&2.15&22.5&\\
\hline
FLAMINGOS (KPNO 4m)&0.2&0.61&$J$&1.25&21.9&\citet{imai07}\\
&&&$K_s$&2.15&21.0&\\
\hline
IRC (AKARI)&0.67&1.46$\times$1.46&$N2$&2.4&21.5&\citet{taka12}\\
&&&$N3$&3.2&21.7&\\
&&&$N4$&4.1&22.1&\\
\cline{3-6}
&&2.34$\times$2.34&$S7$&7.0&19.7&\\
&&&$S9W$&9.0&19.5&\\
&&&$S11$&11.0&19.3&\\
\cline{3-6}
&&2.51$\times$2.39&$L15$&15.0&18.7&\\
&&&$L18W$&18.0&18.7&\\
&&&$L24$&24.0&17.8&
\enddata
\end{deluxetable*}

\section{OBSERVATIONS}  \label{sec:obser}

\begin{figure*}
\epsscale{1.2}
\plotone{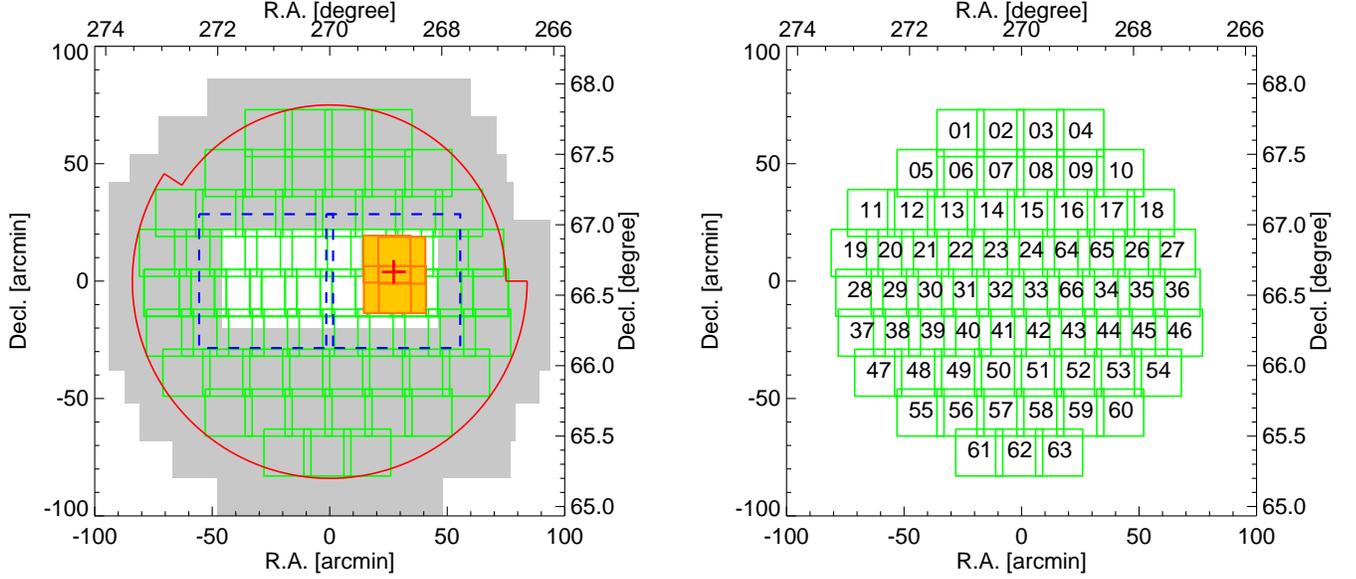}
\caption{Field coverage of the {\it AKARI} NEP multi-wavelength data. The center is the NEP whose coordinates are $\alpha = 18^h00^m00^s,  \delta = +66\degr33\arcmin38\arcsec$. Our $J$ and $H-$band survey (5.1 deg$^2$) is indicated with small empty green squares on both panels (65 subfields). The names of the subfields are marked with numbers on the right panel. The names and central positions for each subfield are listed in Table \ref{t01}. The solid red circles and the two dashed blue squares indicate {\it AKARI} (5.4 deg$^2$) and CFHT MegaCam (2 deg$^2$) coverage, respectively. The gray area represents the $BRI$ imaging survey (4.9 deg$^2$) from Maidanak observatory. The thick red cross indicates the center of NEP-Deep field (0.67 deg$^2$). The $J$ and $K_s$ band coverage from \citet{imai07} (750 arcmin$^2$) are marked with four filled orange squares centered at the NEP-Deep field. 
\label{f02}}
\end{figure*}

\begin{deluxetable*}{cccccc}
\tablecolumns{6}
\tablewidth{0pt}
\tablecaption{
Names of each subfield in Figure \ref{f01} and their central positions. \label{t01}}
\tablehead{
\colhead{Field} & \colhead{R.A.} & \colhead{Decl.} &\colhead{Field} & \colhead{R.A.} & \colhead{Decl.} }
\startdata
FLM01 & 18:04:18.09 & +67:32:00.71 & FLM66 & 17:56:29.34 & +66:28:38.50 \\
FLM02 & 18:01:30.29 & +67:36:38.52 & FLM34 & 17:53:58.87 & +66:28:38.50 \\
FLM03 & 17:58:39.75 & +67:36:38.52 & FLM35 & 17:51:28.40 & +66:28:38.50 \\
FLM04 & 17:55:49.22 & +67:36:38.52 & FLM36 & 17:48:57.93 & +66:28:38.50 \\
FLM05 & 18:07:08.92 & +67:15:36.93 & FLM37 & 18:11:22.13 & +66:11:38.51 \\
FLM06 & 18:04:18.63 & +67:16:45.62 & FLM38 & 18:08:51.66 & +66:11:38.51 \\
FLM07 & 18:01:30.29 & +67:19:38.53 & FLM39 & 18:06:21.19 & +66:11:38.51 \\
FLM08 & 17:58:39.75 & +67:19:38.53 & FLM40 & 18:03:48.46 & +66:10:45.92 \\
FLM09 & 17:55:49.22 & +67:19:38.53 & FLM41 & 18:01:18.11 & +66:10:23.74 \\
FLM10 & 17:52:58.68 & +67:19:38.53 & FLM42 & 17:58:49.78 & +66:11:38.51 \\
FLM11 & 18:10:39.67 & +66:59:10.18 & FLM43 & 17:56:19.31 & +66:11:38.51 \\
FLM12 & 18:07:51.48 & +67:02:38.51 & FLM44 & 17:53:48.84 & +66:11:38.51 \\
FLM13 & 18:05:00.95 & +67:02:38.51 & FLM45 & 17:51:18.37 & +66:11:38.51 \\
FLM14 & 18:02:10.41 & +67:02:38.51 & FLM46 & 17:48:47.90 & +66:11:38.51 \\
FLM15 & 17:59:19.88 & +67:02:38.51 & FLM47 & 18:10:11.92 & +65:54:38.52 \\
FLM16 & 17:56:29.34 & +67:02:38.51 & FLM48 & 18:07:21.38 & +65:54:38.52 \\
FLM17 & 17:53:38.81 & +67:02:38.51 & FLM49 & 18:04:30.85 & +65:54:38.52 \\
FLM18 & 17:50:48.27 & +67:02:38.51 & FLM50 & 18:01:40.31 & +65:54:38.52 \\
FLM19 & 18:11:49.59 & +66:43:31.48 & FLM51 & 17:58:49.78 & +65:54:38.52 \\
FLM20 & 18:09:21.76 & +66:45:38.52 & FLM52 & 17:55:59.25 & +65:54:38.52 \\
FLM21 & 18:06:51.28 & +66:45:38.52 & FLM53 & 17:53:08.72 & +65:54:38.52 \\

FLM22 & 18:04:20.82 & +66:45:38.52 & FLM54 & 17:50:18.18 & +65:54:38.52 \\
FLM23 & 18:01:50.35 & +66:45:38.52 & FLM55 & 18:07:11.35 & +65:37:38.53 \\
FLM24 & 17:59:19.88 & +66:45:38.52 & FLM56 & 18:04:20.82 & +65:37:38.53 \\
FLM64 & 17:56:49.40 & +66:45:38.52 & FLM57 & 18:01:30.29 & +65:37:38.53 \\
FLM65 & 17:54:18.93 & +66:45:38.52 & FLM58 & 17:58:39.75 & +65:37:38.53 \\
FLM26 & 17:51:48.46 & +66:45:38.52 & FLM59 & 17:55:49.22 & +65:37:38.53 \\
FLM27 & 17:49:18.00 & +66:45:38.52 & FLM60 & 17:52:58.68 & +65:37:38.53 \\
FLM28 & 18:11:32.17 & +66:28:38.50 & FLM61 & 18:03:00.56 & +65:20:38.51 \\
FLM29 & 18:09:01.69 & +66:28:38.50 & FLM62 & 18:00:10.03 & +65:20:38.51 \\
FLM30 & 18:06:31.22 & +66:28:38.50 & FLM63 & 17:57:19.50 & +65:20:38.51 \\
FLM31 & 18:03:58.38 & +66:27:09.87 &  & & \\
\enddata
\end{deluxetable*}

The {\it AKARI} NEP field is centered at $\alpha = 18^h00^m00^s,  \delta = +66\degr33\arcmin38\arcsec$ and our NIR data cover the entire {\it AKARI}~'s NEP field area. Figure \ref{f02} (left) shows tiling components of our NIR data (small empty green squares; each of them has a 21$\arcmin$ $\times$ 21$\arcmin$ field of view.) as well as the field coverage of the {\it AKARI} NEP surveys (solid red circles) and its multi-wavelength follow-up observations. The two dashed blue squares show the coverage of the CFHT observation ($u^*g'r'i'z'-$bands) and the gray area is for the Maidanak observation ($BRI-$bands). The four filled orange squares have $J$ and $K_s-$band imaging data introduced in \citet{imai07} covering only the NEP-Deep area. The right panel of Figure \ref{f02} shows the same field coverage as the left one, but shows the field name for each subfield of our NIR data. From FLM01 to FLM66, 65 tiles constitute the NEP NIR area. Three subfields, FLM64 to FLM66, have only $H-$band data from our observation, because this area overlaps with \citet{imai07}'s $J$ and $K_s-$band data. The tile names are used for the target ID (Section \ref{sec:catfm}). The names of each subfield and their central positions are listed in Table \ref{t01}. 

The NIR imaging observation was done with $J$ and $H$ filters from 2008 June 13 to 23 at Kitt Peak National Observatory (KPNO) using the FLoridA Multi-object Imaging Near-ir Grism Observational Spectrometer \citep[FLAMINGOS;][]{elston06}, which is an NIR camera attached to the Cassegrain focal plane of the 2.1m telescope. It uses a Hawaii II 2048x2048 HgCdTe array, divided into four quadrants with 8 amplifiers each. It provides a 21$\arcmin$ $\times$ 21$\arcmin$ field of view, with a pixel scale of 0.606$\arcsec$ pix$^{-1}$. 

\begin{figure*}
\plotone{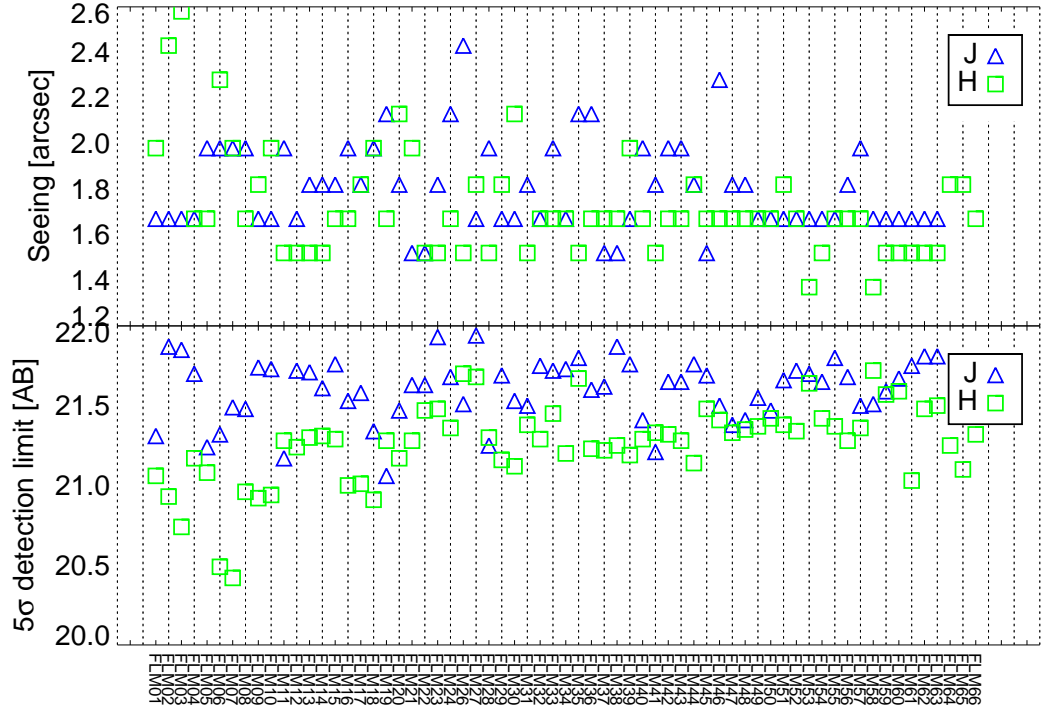}
\caption{
Seeing and 5$\sigma$ detection limits for each subfield. (Triangles for $J$ and squares for $H-$band) 
\label{f03}}
\end{figure*}

\begin{figure*}
\plotone{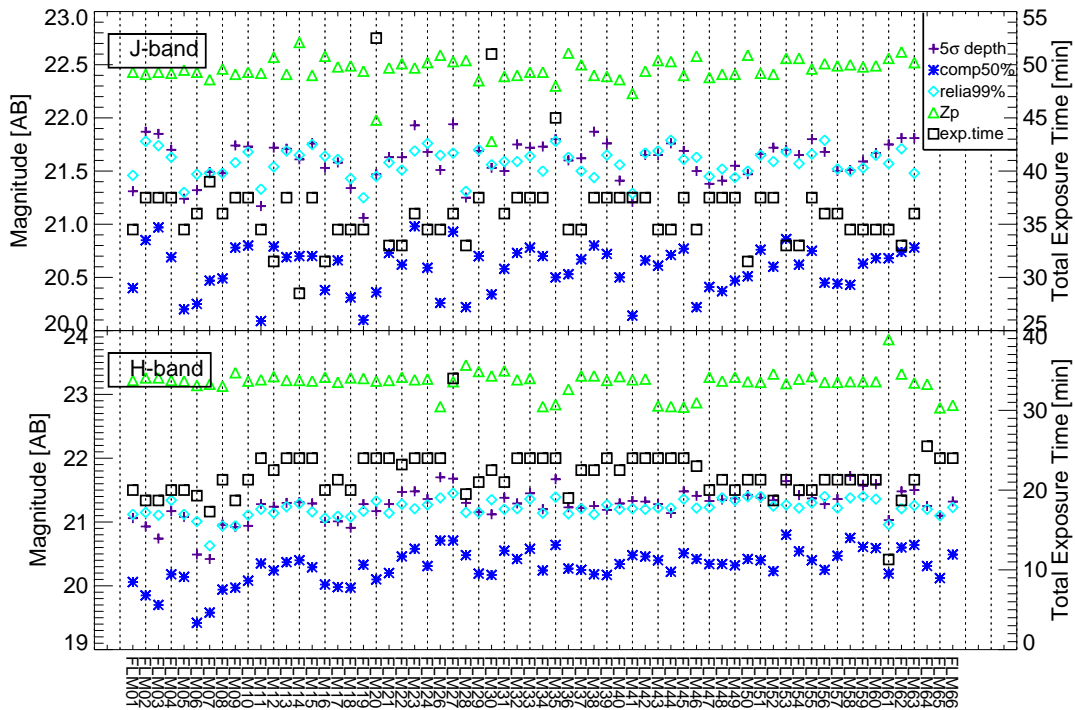}
\caption{Several measures of the image quality and depth of each subfield, such as the 5$\sigma$ detection limit (crosses), the 50$\%$ completeness (asterisks; see Section \ref{sec:compl}), the 99$\%$ reliability (diamonds; see Section \ref{sec:relia}), the photometric zeropoint $Z_p$ (triangles; see Section \ref{sec:calib}), and the total exposure time (squares).  
\label{f04}}
\end{figure*}

The observation was carried out under mostly clear sky conditions. The observations followed a standard sequence which consists of a 5 $\times$ 5 dither pattern with 20$\arcsec$ offsets to make the sky frame for the sky subtraction. Each subfield overlaps with neighboring subfields for regions of width 4$\arcmin$ to avoid the low efficiency parts on the edges of each chip. The total exposure time for each subfield is $\sim$36 minutes for $J-$band and $\sim$21 minutes for $H-$band, and the data reach 5$\sigma$ depths of $\sim$21.6 mag and $\sim$21.3 mag, respectively. We used 1.5 minutes as an exposure time per frame for $J-$band throughout the observations. However, for the $H-$band observations, we employed exposure times between 0.5 and 1.0 minutes per frame to adjust the background level. The seeing values for each subfield were determined as the median values of full width at half maximum (FWHM) of non-saturated point sources with magnitudes between 15 and 19. The seeing values of the stacked $J$ and $H-$band images are determined to be $\sim$1.8$\arcsec$ and $\sim$1.7$\arcsec$, respectively. 
To determine the limiting magnitude, 
we used the aperture size corresponding to a diameter of 3 times the FWHM
following the definition introduced in Section \ref{sec:detec}.
Tables \ref{t02} and \ref{t03} show the observation summary including the total on-source exposure time, the exposure time per frame, the median seeing size, 5$\sigma$ detection limit, and 50$\%$ completeness magnitude (see Section \ref{sec:compl}) for each filter. We plot the values of seeing size and 5$\sigma$ detection limits for all subfields for both filters in Figure \ref{f03}. The figure can be used to check the subfield variance of the values. Also to determine the quality of the images, we show plots of the magnitudes for the  5$\sigma$ detection limits (crosses), 50$\%$ completeness (see Section \ref{sec:compl}), 99$\%$ reliability (see Section \ref{sec:relia}) and photometric zeropoints $Z_p$ (see Section \ref{sec:calib}) for each subfield in Figure \ref{f04}. The total exposure times for each subfield are plotted on the right axis.


\section{DATA REDUCTION} \label{sec:datar}
\subsection{Preprocessing}  \label{sec:prepr}

For the data reduction, we first carried out i) linearity correction, ii) dark subtraction and flat fielding, iii) distortion correction, iv) sky subtraction, v) bad pixel and cosmic ray masking, and vi) image alignment and combining. All these steps are performed using IRAF\footnote{IRAF is distributed by the National Optical Astronomy Observatory, which is operated by the Association of Universities for Research in Astronomy (AURA) under cooperative agreement with the National Science Foundation.}.

To correct the nonlinearity of the NIR detector, we used the IRAF \texttt{irlincor} task. For FLAMINGOS, the linearity correction can be expressed as a third-order polynomial\footnote{ http://www.noao.edu/kpno/manuals/flmn/flmn.user.html}:
\begin{displaymath}    
   y = x \times ( a + b \times (x / 32767) + c \times (x / 32767)^2 ),
\end{displaymath}
where $x$ is the raw signal, $y$ is the linearity corrected signal in ADU, and  the coefficients are $a = 1.00425, b = -0.03323, c = 0.04489$.

We used a standard method for dark subtraction and flat fielding with the IRAF \texttt{noao.imred.ccdred} package. Master dark frames were created by combining dark frame image for each night and exposure time. We subtracted the master dark image from each science image. These dark subtracted science images, typically 250 frames with a background count of 3000 DNs, were combined in median after scaling them to a reference image by the mode pixel value to construct master sky flat images for each night. 
We compared this master sky flat image with flat frames taken from dome dimming lights and found that the master sky flat can remove ring structures in the center of science images more efficiently. The flat fielding was done by dividing dark subtracted images by the normalized master sky flat image for each night. We removed satellite tracks on the images using the IRAF \texttt{satzap} task. 

The distortion correction\footnote{http://flamingos.astro.ufl.edu/DistortionFiles/index.html} was performed using a distortion map constructed from pinhole data. We used the IRAF \texttt{geotran} task for the correction. Among the various distortion maps, we used the most resent one, KP2m04.map, obtained in July 2003. 
The distortion map doubles the number of pixels in each axis and sets the pixel scale to 0.303$\arcsec$ pix$^{-1}$. We compared the astrometric accuracy before and after the distortion correction (see Section \ref{sec:astrm} for more details about the astrometry). We used the FLM14 $J-$band field as a test field. 
The astrometric accuracy was improved by up to $\sim$5$\%$ after the distortion correction.

To subtract the sky, mask bad pixels and cosmic rays, and align and combine the sky subtracted individual images, we used the Experimental Deep Infrared Mosaicing Software \citep[\texttt{XDIMSUM};][]{stan95} package of IRAF. Six adjacent frames were used for making the sky image to subtract the sky. 
The mask images were created using the master dark images and master sky flat images for each night, excluding bad pixels which have unusual values compared with the normal distribution of the pixel values. Cosmic rays are removed with a 3$\sigma$ clipping method while combining the images. 

\subsection{Astrometry}  \label{sec:astrm}

\begin{figure*}
\epsscale{1.15}
\plotone{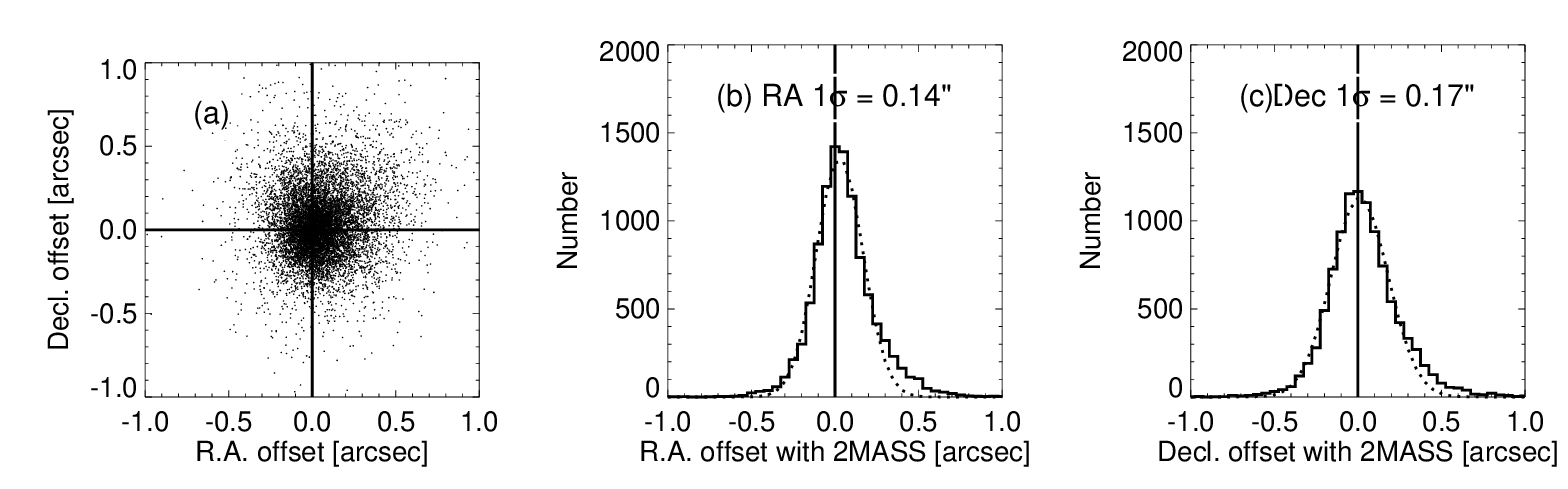}
\caption{(a): Distribution of the differences in position between our sources and the corresponding 2MASS sources in R.A. and Decl. directions. (b) and (c): Histograms of the distributions in R.A. and Decl. directions, respectively. The dotted lines indicate the best fitted Gaussian function for the histograms. From the Gaussian function, the 1$\sigma$ astrometry errors are 0.14$\arcsec$ for R.A. and 0.17$\arcsec$ for Decl.. 
\label{f05}}
\end{figure*}

We employed SCAMP \citep{bert06} which finds astrometric solution by cross matching between positions of input catalog created from a science image without astrometry calibration and a reference catalog containing a set of accurate R.A. and Decl. values. The Two Micron All Sky Survey Point Source Catalog (2MASS PSC) was used as the reference catalog. The astrometric solution was derived for each subfield and each filter separately. We tried to find the best parameter value for the degree of the polynomial for distortion correction between 3, 4, 5, and 6 and found that a $5^{th}$-order polynomial gave the best astrometric accuracy among other options. Therefore we set the degree of the polynomial to 5. 

To check the astrometric accuracy, 2MASS point sources with 1$\sigma$ position uncertainty less than 0.2$\arcsec$ are compared with our catalog. Figure \ref{f05} (a) shows the distributions of the positional differences between our sources and the corresponding 2MASS sources in R.A. and Decl. directions, and (b) and (c) show the histogram of the position differences in the two directions, respectively. We fit the histograms with a Gaussian function (dotted lines) and take the 1$\sigma$  value of the best fitted Gaussian function as the 1$\sigma$ astrometric errors in R.A. and Decl. direction. 
Here, the 1$\sigma$ errors are around 0.16$\arcsec$ for both directions. 
Combining the rms scatter in both directions in quadrature, we get the absolute positional accuracy of 0.17$\arcsec$.
Considering that the astrometry error of the 2MASS sources is about 0.2$\arcsec$, our astrometric accuracy is quite good, and most of the uncertainty in the astrometry seems to originate from the 2MASS catalog. 

\subsection{Photometric Calibration} \label{sec:calib}

\begin{figure*}
\epsscale{1.2}
\plotone{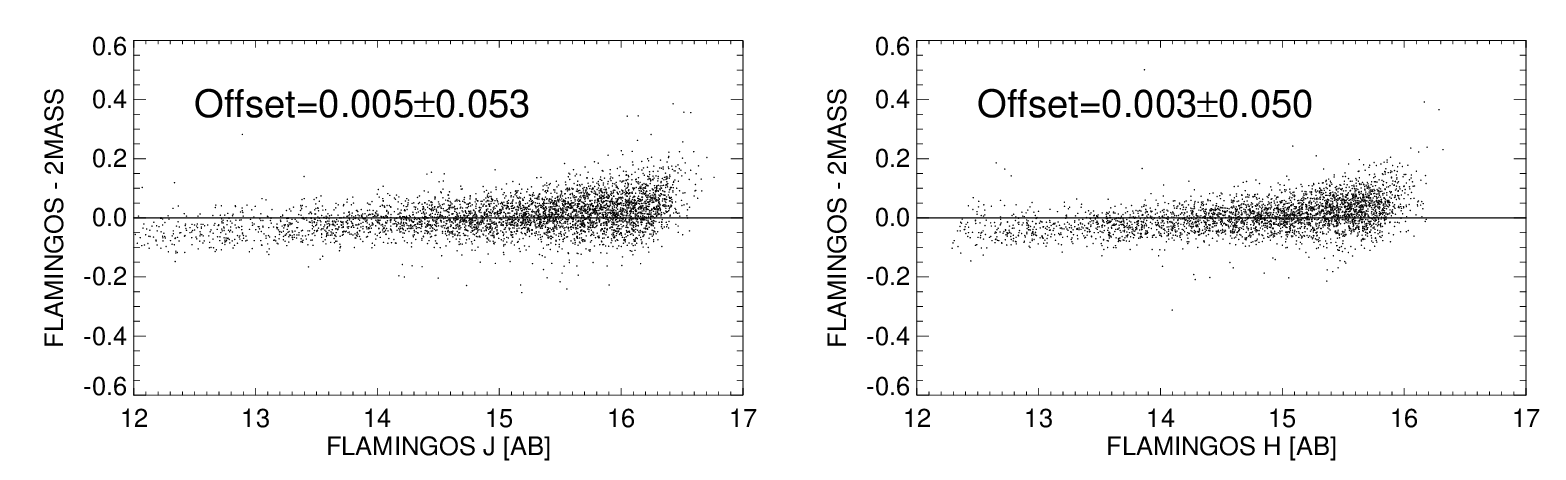}
\caption{Photometric difference between our sources and the corresponding 2MASS sources as a function of magnitude to check the photometric accuracy. Although the systematic offset is very small, the bright sources show average offsets of $\sim$0.05 for both filters. The scatter of the magnitude difference is larger for fainter sources, which may be caused by the photometric uncertainties for the sources. The dispersion of the magnitude difference are $\sim$0.05 mag. These values are corresponding with the zeropoint uncertainty.
\label{f06}}
\end{figure*}

For the photometric calibration, we also used stellar objects in the 2MASS PSC. 
We restricted the photometric reference stars to have the color $-0.9 < J_{2MASS} - K_{2MASS} < 0.1$ (0 $< J_{2MASS} - K_{2MASS} < $ 1 for the Vega system; see Section \ref{sec:catfm}), photometric error less than 0.05 mag, and magnitude fainter than 11.9 mag (11 mag for the Vega system). 
For each field, we derived the difference between the 2MASS and instrumental magnitudes of each source in the field and computed the average value of the difference to define the zeropoint ($Z_p$). The zeropoint error was derived from the standard deviation of the magnitude offset. 
The average zeropoint error is $\sim$0.05 mag for both $J$ and $H-$band.
The $Z_p$ for each subfield is listed in Table \ref{t03} for the purpose of checking the depth of each field. 
The $Z_p$ is used for converting the $DN$ per exposure time to magnitudes, i.e. $M=Z_p - 2.5\log$ $(DN/sec)$. 
Figure \ref{f04} shows the distribution of several depth indicators for each field. The depths are checked using the 5$\sigma$ detection limit, the 50$\%$ completeness (see Section \ref{sec:compl}), the 99$\%$ reliability (see Section \ref{sec:relia}), and the $Z_p$ (triangles), as well as the total exposure time.

To check the photometric accuracy, we compare the magnitudes of 2MASS sources with our sources. 
Figure \ref{f06} shows the magnitude differences between our sources and the corresponding 2MASS sources as a function of magnitude for $J$ (left panel) and $H$ (right panel) band.
We used the 2MASS point sources of same color and photometric error properties that we used for the $Z_p$ calculation.
The magnitude offsets show very small systematic differences (0.002 for $J-$band and 0.001 for $H-$band). 
For the brightest sources ($J$ and $H \lesssim 13.5$), however, they show offsets of $\sim$0.05 for both filters. Since the halo region near a very bright star has a dark background, it causes the flux of the bright star to be overestimated. The dark halos are induced by the sky subtraction during preprocessing because bright stars in the dithered images overestimate the sky values. The scatter of the magnitude difference is larger for fainter sources, which may be caused by the photometric uncertainties for the sources. We calculated the variances of the offsets for both filters using the standard deviation of the offsets. They have values of $\sim$0.05, which coincide with the zeropoint uncertainties. Therefore the magnitudes of our catalog are accurate to $\sim$0.05 mag.

\section{CATALOGS PROPERTIES}\label{sec:catpr}

\subsection{Object Detection and Photometry} \label{sec:detec}

To detect sources and conduct photometry, we employed SExtractor \citep[SE;][]{bert96}. 

We found optimal parameters for SE to obtain high completeness while not sacrificing reliability of the detection (Sections \ref{sec:compl} and \ref{sec:relia}) for all  magnitude bins. 
To do so, we made artificial objects using the IRAF \texttt{noao.artdata} package and they are placed randomly at the background regions of the images. 
We performed detections and photometry with various parameter settings.
After many trials with varying SE parameters, we chose the following set of detection parameters for SE: DETECT\_MINAREA=20, DETECT\_THRESH=1.1, DEBLEND\_NTHRESH=64, DEBLEND\_MINCONT=0.005, BACK\_SIZE=500, BACK\_FILTERSIZE=2. The adopted detection parameters corresponds to $S/N=5$. 

We detected 208,020 sources and 203,832 sources for $J$ and $H-$band, respectively, at the magnitude limits brighter the 99$\%$ reliability (see Section \ref{sec:relia}). 
Among these objects, 28.3$\%$ in $J$ and 30.8$\%$ in $H$ are detected only in a single filter image. 
Most of the sources detected only in a single filter image are those at near the magnitude limits, suggesting that they are missed in an image in the other band due to the detection limit. 
The reason for the percentage of sources detected only in $J-$band being slightly higher is the deep $J-$band catalog from \citet{imai07}. If a bright object of $J$ $<$ 19 mag or $H$ $<$ 20 mag has no counterpart in the other band, we find that they are a crosstalk or cosmic ray. 
We derived auto magnitudes with Kron-like elliptical aperture \citep{kron80}, which are assumed to be total magnitudes. 
For the aperture magnitudes, we set the aperture diameter size to 3 times the FWHM. 
We found a diameter of 3 times the FWHM to be an optimal diameter 
which can contain the majority of the flux from an object 
but is not affected by the local fluctuation in the sky background.
This has been checked by growth curves of objects with various magnitudes.
The magnitude errors are computed by combining the standard SE errors based on Poisson statistics and the zeropoint errors. At the bright end, the photometry is free from  saturation above $J \sim$10 mag and $H \sim$11 mag.

Since the stacking of the dithered frames with sub-pixels during preprocessing causes poor measurement of pixel-by-pixel rms for the  correlated neighboring pixels, the errors from SE, which only consider the Poisson noise, can be underestimated. 
To improve the accuracy of the estimated errors, we adopted the method in \citet{gaw06} and \citet{jeon10}, calculating the noise rate induced by the pixel-by-pixel correlated noise. We place $\sim$3000 apertures of a given size randomly located in the background area. The aperture has an area of $n_{pix}$ and an effective diameter of $N = n_{pix}^{1/2}$. The background rms fluctuations are derived from a Gaussian curve fit to the histogram of fluxes inside each aperture. The noise calculated from the rms fluctuations for a given effective diameter is expressed as $\sigma_N \propto N^\beta$. If the noise is only dependent on the Poisson noise, it is independent from one pixel to another. Therefore for this case, $\sigma_N \propto N$. However, for the adjoining pixels, the sky noise has $\beta > 1$. 
For our data, we find $\sigma_{N} \propto N^{1.4}$. For an aperture with 5.4$\arcsec$ diameter (3 times the typical FWHM of
our images), there are about 250 pixels (or $N \simeq 16$). Therefore, the photometric errors of faint objects ($J \gtrsim 18$ mag) may be underestimated by a maximum factor of 3. For the brighter objects  ($J \lesssim 16$ mag), the photometric errors are not affected by this effect, since their errors are dominated by noises from the source. 

The detection and photometry of the sources are carried out for each subfield image, rather than a single mosaic of all the subfields, since there is a variation in the weather condition among subfields. Catalogs from individual subfields are then merged together to form a master catalog for each filter. However, since each subfield image overlaps with a neighboring images, catalogs created from each subfield image contain duplicate entries in the overlap region. 
To deal with the duplicate entries during the merging process of all the subfield catalogs, we set the criterion for choosing the entry with the smallest FLAGS values from SE and smallest photometric errors for multiple entries with positional offsets less than 1.5$\arcsec$.
Only the object chosen with the above criterion, and its SE output values derived from the chosen subfield are included in the final catalog. 

When we merged the $J$ and $H-$band catalogs, we used a 1.5$\arcsec$ radius. Every row, regardless of whether they have information in both bands, is included in the final catalog. Regarding the stellarity from SE, we chose a single stellarity value for the final catalog where the photometric error of the source is smaller between the two filters. 

\subsection{Crosstalk Flagging} \label{sec:crstk}
The FLAMINGOS chip exhibits crosstalk artifacts between amplifiers on both sides of a bright source\footnote{http://flamingos.astro.ufl.edu/DetQE/index.html}, where the crosstalk signals can be identified as doughnut-shaped signals at every 128 pixels, which corresponds to the width of one amplifier, from the position of a bright source. 
The strength of the crosstalk signal is stronger for a brighter source, and it decreases as the distance from the bright source increases.
We examined the magnitude of the source that generates crosstalks and the positions of the crosstalks. 
The crosstalks start to be seen on both sides of a bright source at a magnitude of $J \sim 15.0$ mag and $H \sim 15.5$ mag in the stacked image.
For sources that are about 3 magnitudes  brighter than that limit, the crosstalks continue to the end of a quadrant. 
The bright sources create the crosstalk signal at distances of multiples of 256 pixels in our stacked image whose pixel sizes are a half of the pixel size of the raw image. 
The direction in which the crosstalks appeared depends on the readout direction and where the object is located, but it is predictable.
Using these properties, we predicted the positions of the crosstalks and included the flag information for objects at these points in our merged catalog that the object is a potential crosstalk signal or its photometry is affected by the crosstalk.

\subsection{Data from \citet{imai07}}  \label{sec:imais}
The $J$ and $K_s-$band imaging was conducted in 2004, covering 750 arcmin$^2$ near the center of NEP-Deep by \citet{imai07}, where we avoided duplicate observation in $J-$band. 
To construct a continuous area coverage of the $J-$band imaging in the NEP field, we included the $J-$band images from \citet{imai07} when creating a catalog of $J-$band sources. 
For consistency in the final catalog, we conducted new detections and photometry using our own method (see Section \ref{sec:detec}). Their flux calibration was done using standard stars listed in \citet{pers98}, resulting in uncertainties in the flux calibration up to 0.1 mag. We were able to reduce the errors to less than 0.05 mag using our flux calibration method (see Section \ref{sec:calib}). We added this catalog to the final catalog to fill the gap in the NEP field from our $J-$band survey. The average depths of images of \citet{imai07} are $\sim$0.3 mag deeper than those of our data.

\subsection{Star-Galaxy Separation} \label{sec:galcut}

\begin{figure*}
\plotone{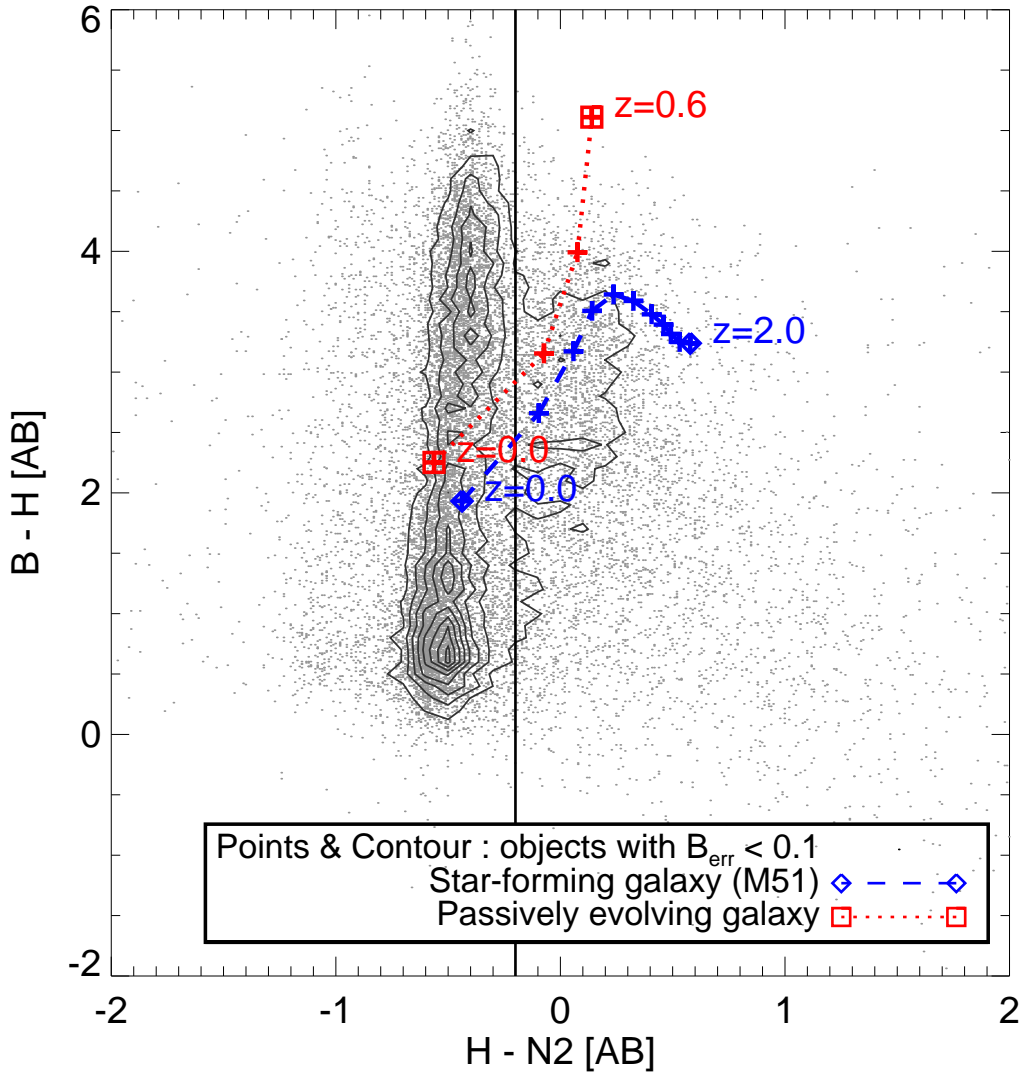}
\caption{ $B-H$ versus $H-N2$ color-color diagram and the density contours of sources detected in $B$, $H$, and $N2-$band. The points and the contours are made with objects that have $B-$band magnitude errors less than 0.1 mag. The dashed line is a redshift track of a star-forming galaxy (M51) and the dotted line is for a passively evolving galaxy made with the \citet{bruz03} model of a passively evolving 5 Gyr-old galaxy with a spontaneous burst, a metallicity Z=0.02, and  the Salpeter IMF. We use the vertical line as a boundary that distinguishes stellar sources against galaxies.  
\label{f07}}
\end{figure*}

We used the same method as \citet{jeon10} to separate galaxies from stars. Since most stars have similar slopes at wavelengths greater than 1$\micron$ on their SEDs, NIR colors of stars lie in narrow range while those of distant galaxies are more diverse due to redshifted SED. Figure \ref{f07} shows the $B-H$ versus $H-N2$ color-color diagram. The stellar locus can clearly be seen by the clouded gray points and overlapping contour ($H-N2 \sim$ 0.5). Also the redshift tracks of two types of galaxies are shown to check the location of galaxies. The dashed line is a redshift track of a star-forming galaxy (M51) and the dotted line is that of a passively evolving 5 Gyr-old galaxy with a spontaneous burst, a metallicity Z=0.02, and the Salpeter initial mass function (IMF) created with the \citet{bruz03} model. 
As indicated in Figure \ref{f07}, we find that the galaxies and stars can be separated effectively with a color cut at $H - N2 = -0.2$ mag, where objects with $H - N2 < -0.2$ are mostly stars. However, galaxies at z $\sim 0$ also have $H-N2 < -0.2$ mag, and there are also stars with peculiar colors with $H - N2 > -0.2$. 
To deal with such case, we also use the stellarity parameter from SE
of objects in the optical catalog \citep{jeon10,hwan07} since the optical images offer better spatial resolution ($\sim 1\arcsec$). 
The SE stellarity parameter has a value between 1 (point source) and 0 (extended source). For the galaxies with $H-N2 < -0.2$  mag, they can be selected using the criterion stellarity $\leq$ 0.8. 
Likewise, point sources with stellarity $>$ 0.8 with $H-N2 > -0.2$ mag are classified as stars. 
This method was applied only to objects brighter than the 50$\%$ completeness magnitude, which corresponds to $R <$ 20 mag or $r <$ 22 (see Figure 6 in \citet{jeon10} and Figure 2 in \citet{hwan07}), since the 50$\%$ completeness magnitude is the limit for which the stellarity parameter can be used to separate the extended sources from the point sources.
Near the detection limit, all objects fainter than the magnitude of 50$\%$ completeness are considered as galaxies. 
 
\subsection{Catalog Format}  \label{sec:catfm}

\begin{figure*}
\plotone{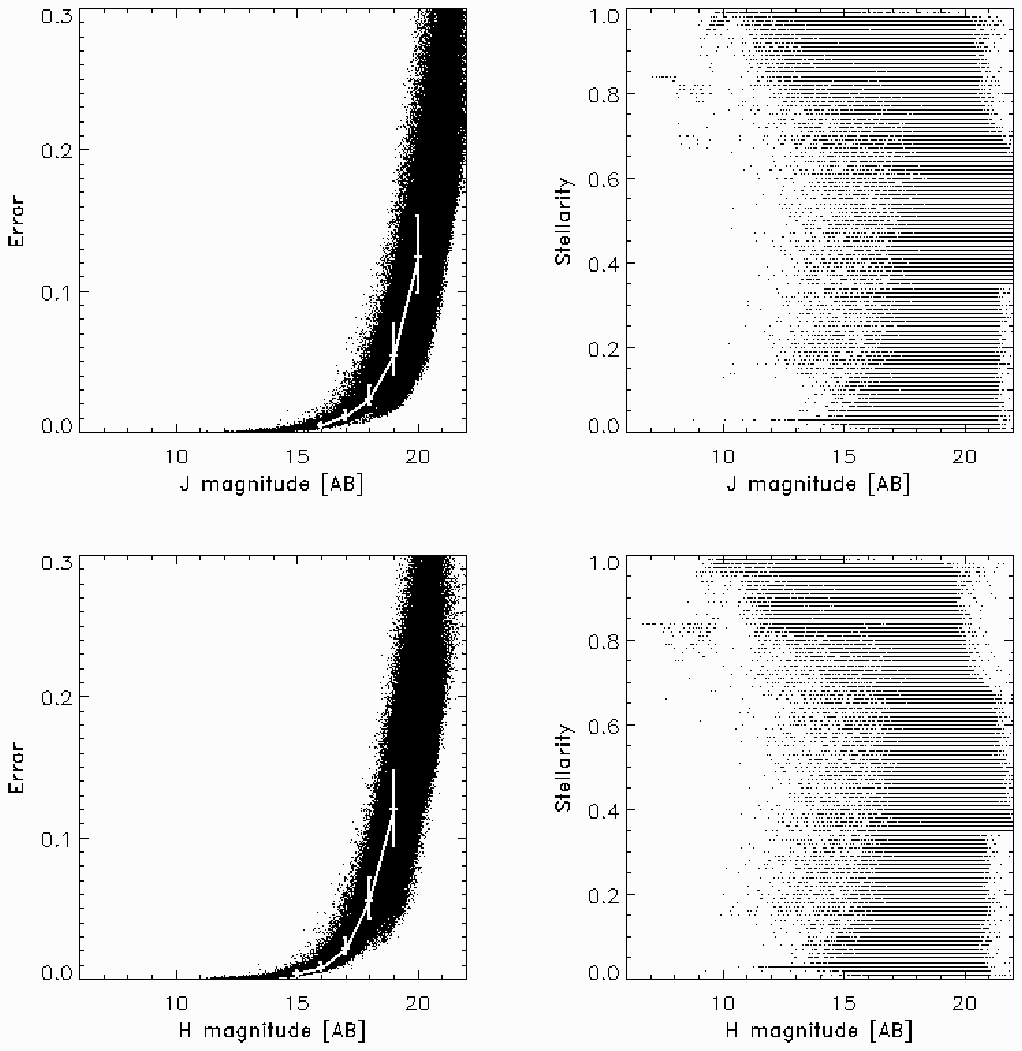}
\caption{The photometric error distribution and the stellarity distribution \label{f08}}
\end{figure*}

Table \ref{t04} shows a part of the $J$ and $H-$band merged catalog. The catalog contains objects whose detection reliability is higher than 99$\%$. The magnitude cut was computed independently for each field in each filter (see Section \ref{sec:relia}). The catalog contains 15 columns; identification number (ID), the right ascension (R.A.) and declination (Decl.), $J$ and $H-$band magnitudes and errors, stellarity, galactic extinction values for each filter, and the information about the crosstalk. The object IDs consist of their field name and number in the order of their R.A. for each field. For both $J$ and $H-$band, we provide two kinds of magnitudes; auto magnitude ($J_{AUTO}$ and $H_{AUTO}$) using Kron-like elliptical aperture and aperture magnitude ($J_{APER}$ and $H_{APER}$) within a circular aperture with a diameter of three times FWHM. The magnitude errors are computed from the square root sum of the Poisson noise from the source and the zeropoint error from the photometric calibration. The zeropoint errors have values less than 0.05 mag. The distributions of the errors are shown in Figure \ref{f08}. The large dispersions of the errors are induced by the different depths of each field. The stellarities have values between 0 (extended source) and 1 (point source), and the distributions of the stellarities are also shown in Figure \ref{f08}. The galactic extinction values are from the extinction map of \citet{schl98}. Since the color excess $E(B-V)$ has values of 0.034 -- 0.058, $A_J\sim0.04$ and $A_H\sim0.03$, which are both smaller than the photometric zeropoint errors. Note that the magnitudes in the final catalog are not corrected for the galactic extinction. The final column contains information regarding the crosstalks; the {\it crt} means that the object is located on a crosstalk position. We used a dummy value of 99.999 for non-detections. If the object has only single filter data, we used {\it NaN} for the other filter information.

Using the filter response curve data of FLAMINGOS\footnote{http://flamingos.astro.ufl.edu/Filter\_Info/index.html} and the Vega spectrum from \citet{bohl04}, we computed the conversion factor between AB and Vega magnitude. The conversion formulae from Vega to AB magnitudes are as follows:
\begin{displaymath}J(Vega) = J(AB) - 0.926\end{displaymath}
\begin{displaymath}H(Vega) = H(AB) - 1.367\end{displaymath}
\begin{displaymath}K_s(Vega) = K_s(AB) - 1.798\end{displaymath}
We apply this method to the 2MASS filter system ($J_{2MASS}$ and $H_{2MASS}$) and found that they have values similar to those  of the FLAMINGOS filter system, 0.910 for $J_{2MASS}$ and 1.383 for $H_{2MASS}$.

\subsection{Completeness} \label{sec:compl}

\begin{figure*}
\plotone{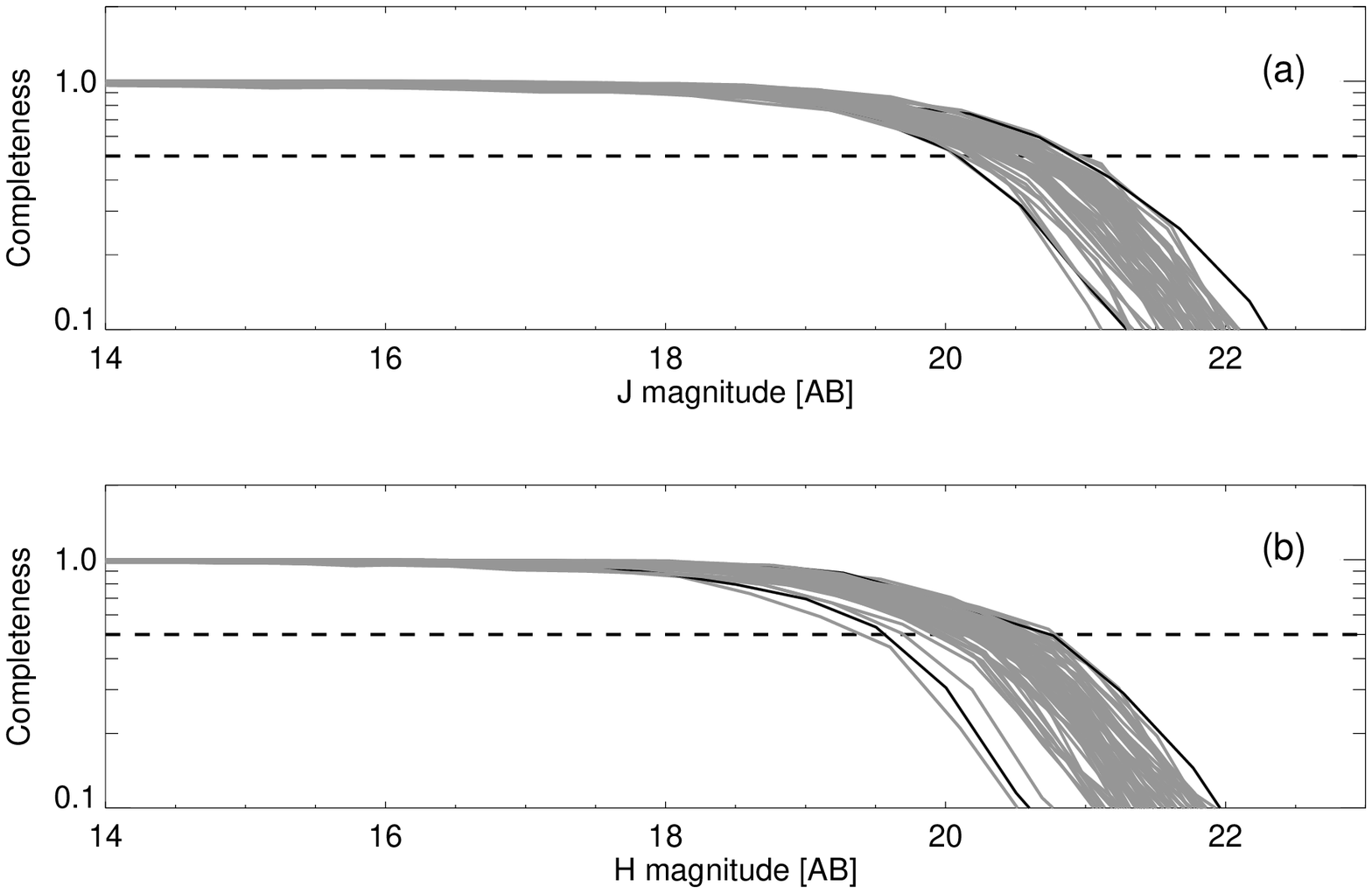}
\caption{Completeness (the ratio of the number of detected artificial sources to the number of input objects in a given magnitude bin) for $J$ and $H-$band. The gray solid lines are the completeness of all subfields from the simulation method, especially for the deepest fields (FLM27 in $J$ and FLM58 in $H$; the upper black solid lines) and the shallowest fields (FLM19 in $J$ and FLM07 in $H$; the lower black solid lines) in the 5$\sigma$ detection limits. The horizontal dashed line is for 50$\%$ completeness cut. The magnitudes of 50$\%$ completeness for each subfield are shown in Table \ref{t03}.
\label{f09}}
\end{figure*}

To estimate the completeness for each field, we used artificial objects from the \texttt{noao.artdata} package of IRAF. The artificial objects are created with magnitude between 11 and 25 and they are placed randomly in the background regions of the images. The number of point sources and extended sources created are 100 each. Among the extended sources,  40$\%$ are elliptical galaxies while the others are disk galaxies, with a minimum redshift of z $\sim 0.01$. The angular sizes of the galaxies, which are equivalent to the half-flux diameters of the model objects, are set to be the FWHM values for each image. We performed detections and photometry with the same parameter settings of SE which were applied for the actual images. We define the completeness as the ratio of the number of detected artificial sources to the number of input objects in a given magnitude bin. Figure \ref{f09} represents the completeness of $J$ and $H-$band data as a function of magnitude. The gray lines represent the completeness for each subfield. The thick, black lines are for the deepest (FLM27 in $J$ and FLM58 in $H$) and the shallowest (FLM19 in $J$ and FLM07 in $H$) fields according to the 5$\sigma$ detection limits. In Figure \ref{f04}, we compared the magnitude of 50$\%$ completeness (asterisks) for each field with the 5$\sigma$ detection limit, the 99$\%$ reliability (see Section \ref{sec:relia}), the $Z_p$, and the total exposure time. The magnitudes of 50$\%$ completeness for each field are listed in Table \ref{t03}. We can see that the magnitudes of 50$\%$ completeness are $\sim$1 mag brighter than the 5$\sigma$ detection limits.

\subsection{Reliability} \label{sec:relia}

\begin{figure*}
\plotone{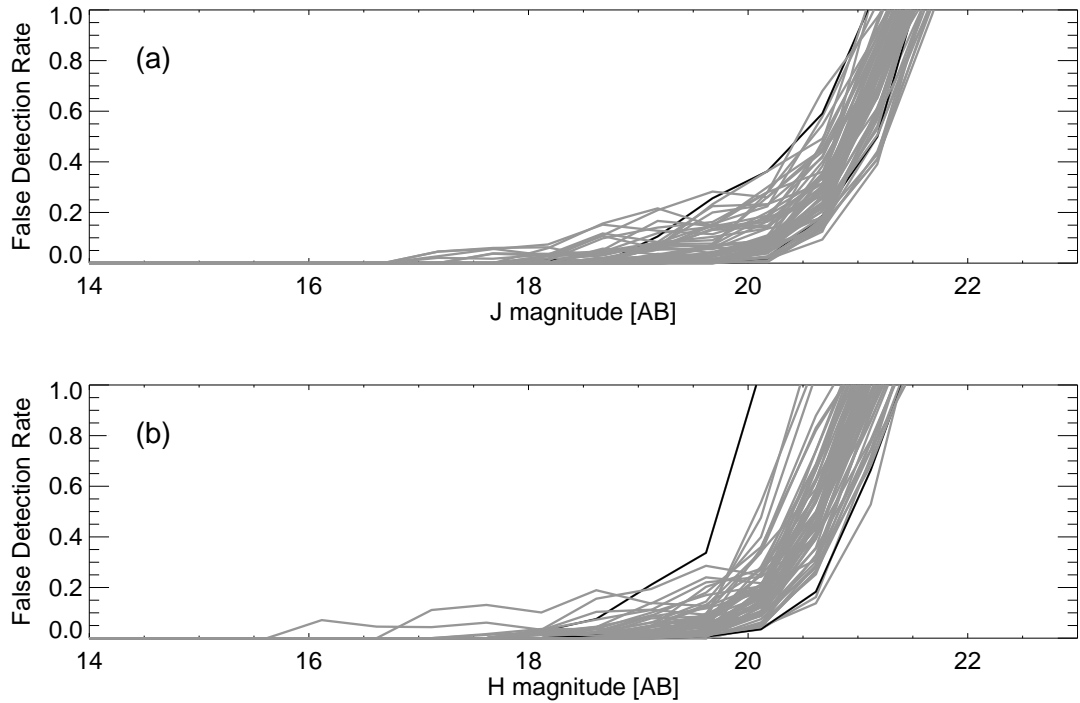}
\caption{ False detection rate (the ratio of the number of objects detected in the negative image to the number of objects detected in the original image for each magnitude bin) of all subfields in $J$, and $H-$band. The black solid lines are for the deepest or the shallowest fields as same as Figure \ref{f09}. The magnitudes of false detection rate 1$\%$ for each field are shown in Table \ref{t03}. 
\label{f10}}
\end{figure*}

To examine the reliability of our source detection, we used the false detection rate for each field and filter. The false detection rate is defined as the ratio of the number of objects detected in a negative image to the number of objects detected in the original image in a given magnitude bin. The negative images were created by multiplying $-1$ to the original images. 
If an image has only noise, its false detection rate would be 1. 
We performed detections and photometry on the negative image with the same parameter settings of SE which were applied for the actual images. 
Figure \ref{f10} represents the false detection rates of $J$ and $H-$band data as a function of magnitude. 
The gray lines represent the false detection rates for each subfield. The thick, black lines are for the deepest (FLM27 in $J$ and FLM58 in $H$) and the shallowest (FLM19 in $J$ and FLM07 in $H$) fields according to the 5$\sigma$ detection limits. 
In Figure \ref{f04}, we compared the magnitude of 1$\%$ false detection rates (equivalent to 99$\%$ reliability) (diamonds) for each field with the 5$\sigma$ detection limit, 50$\%$ completeness, the $Z_p$, and the total exposure time. The magnitude of 99$\%$ reliability for each field are presented in Table \ref{t03}. We can see that the magnitudes of 99$\%$ reliability are similar with the 5$\sigma$ detection limits.

\section{PROPERTIES OF THE DATA}\label{sec:datap}
 
\subsection{Source Number Counts} \label{sec:galnum}

\begin{figure*}
\epsscale{0.9}
\plotone{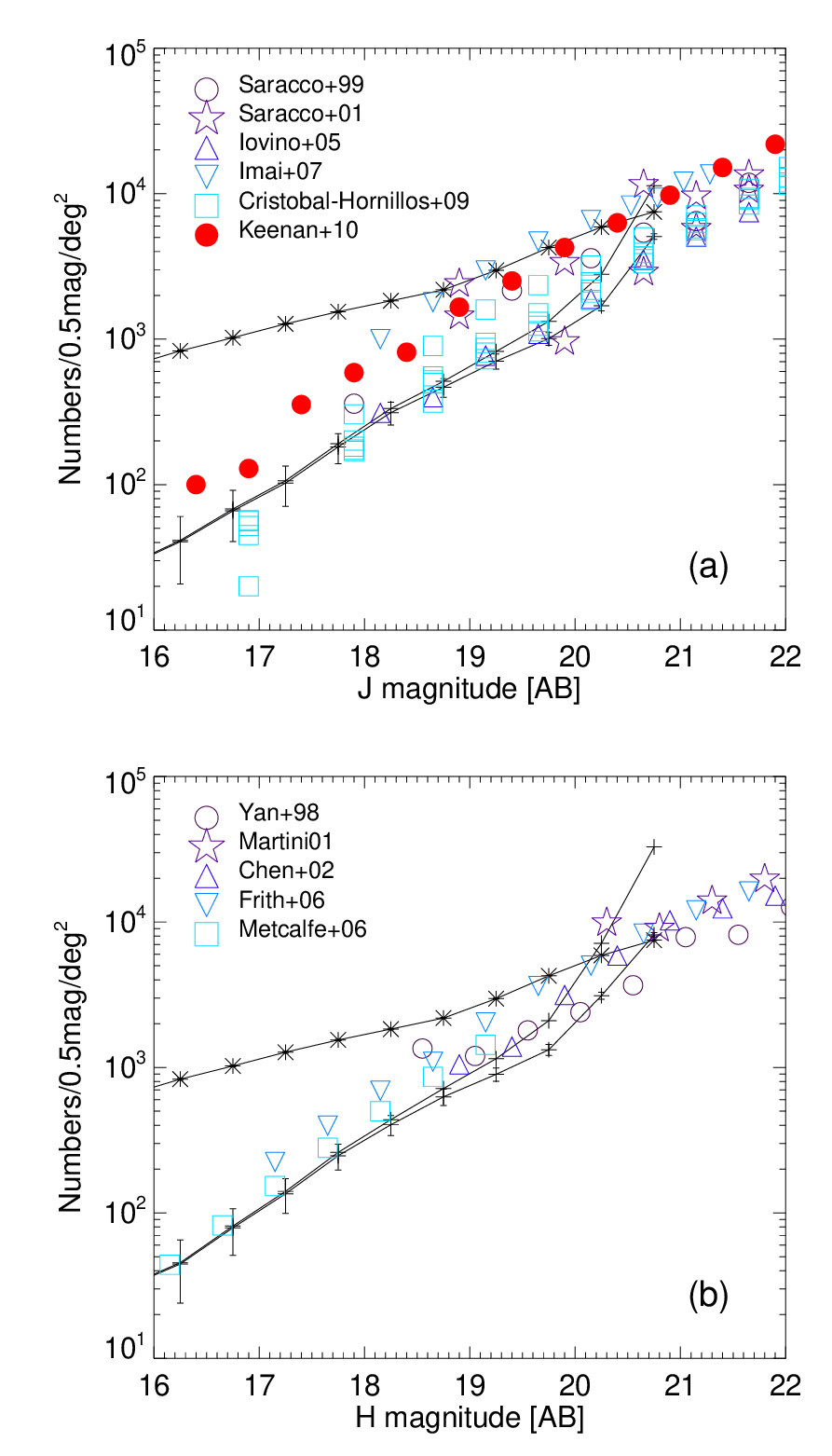}
\caption{ Galaxy (two solid lines with crosses) and stellar (solid lines with asterisks) number counts of $J$ and $H-$band. The error estimates of number counts are from the Poisson errors. The lower solid line is for galaxy number count without completeness correction, and the upper solid line is from the galaxy number count corrected using the completeness. We compare our number counts with the number counts in the literature (other symbols). The galaxy and stellar number count data are provided in Table \ref{t05}.  
\label{f11}}
\end{figure*}

\begin{deluxetable*}{cccccccccc}
\tablecolumns{10}
\tablewidth{0pc}
\tablecaption{ 
Galaxy and stellar number counts. \label{t05}}
\tablehead{
\colhead{} &  \multicolumn{4}{c}{$J-$band} &\colhead{}&  
\multicolumn{4}{c}{$H-$band} \\
\cline{2-5} \cline{7-10}   \\
\colhead{mag} & \colhead{Galaxy}& \colhead{} & \colhead{Corrected} &\colhead{Star} & \colhead{} &
\colhead{Galaxy}& \colhead{} & \colhead{Corrected} &\colhead{Star}\\
\colhead{} & \colhead{Number}& \colhead{Error} & \colhead{Number} & \colhead{Number}& \colhead{} &
\colhead{Number}& \colhead{Error} & \colhead{Number}& \colhead{Number}} 
\startdata

  16.25&    4.05e+01&    1.98e+01&    9.94e-01&    8.31e+02&&    4.45e+01&    2.05e+01&    9.94e-01&    8.31e+02\\
  16.75&    6.59e+01&    2.54e+01&    9.85e-01&    1.02e+03&&    7.89e+01&    2.78e+01&    9.87e-01&    1.02e+03\\
  17.25&    1.02e+02&    3.15e+01&    9.78e-01&    1.27e+03&&    1.35e+02&    3.64e+01&    9.78e-01&    1.27e+03\\
  17.75&    1.82e+02&    4.23e+01&    9.67e-01&    1.54e+03&&    2.47e+02&    4.98e+01&    9.63e-01&    1.54e+03\\
  18.25&    3.13e+02&    5.59e+01&    9.52e-01&    1.84e+03&&    4.04e+02&    6.37e+01&    9.37e-01&    1.84e+03\\
  18.75&    4.67e+02&    6.84e+01&    9.25e-01&    2.18e+03&&    6.28e+02&    7.99e+01&    8.90e-01&    2.18e+03\\
  19.25&    7.06e+02&    8.45e+01&    8.71e-01&    2.97e+03&&    8.96e+02&    9.54e+01&    7.94e-01&    2.97e+03\\
  19.75&    1.01e+03&    1.01e+02&    7.73e-01&    4.27e+03&&    1.31e+03&    1.15e+02&    6.42e-01&    4.27e+03\\
  20.25&    1.64e+03&    1.27e+02&    6.17e-01&    5.89e+03&&    3.00e+03&    1.70e+02&    4.42e-01&    5.89e+03\\
  20.75&    4.96e+03&    2.23e+02&    4.54e-01&    7.50e+03&&    8.11e+03&    2.87e+02&    2.52e-01&    7.50e+03\\

\enddata
\tablecomments{
Units of the galaxy and stellar number densities and their errors are numbers 0.5 mag$^{-1}$ deg$^{-2}$. 
}
\end{deluxetable*}

\begin{figure*}
\epsscale{1}
\plotone{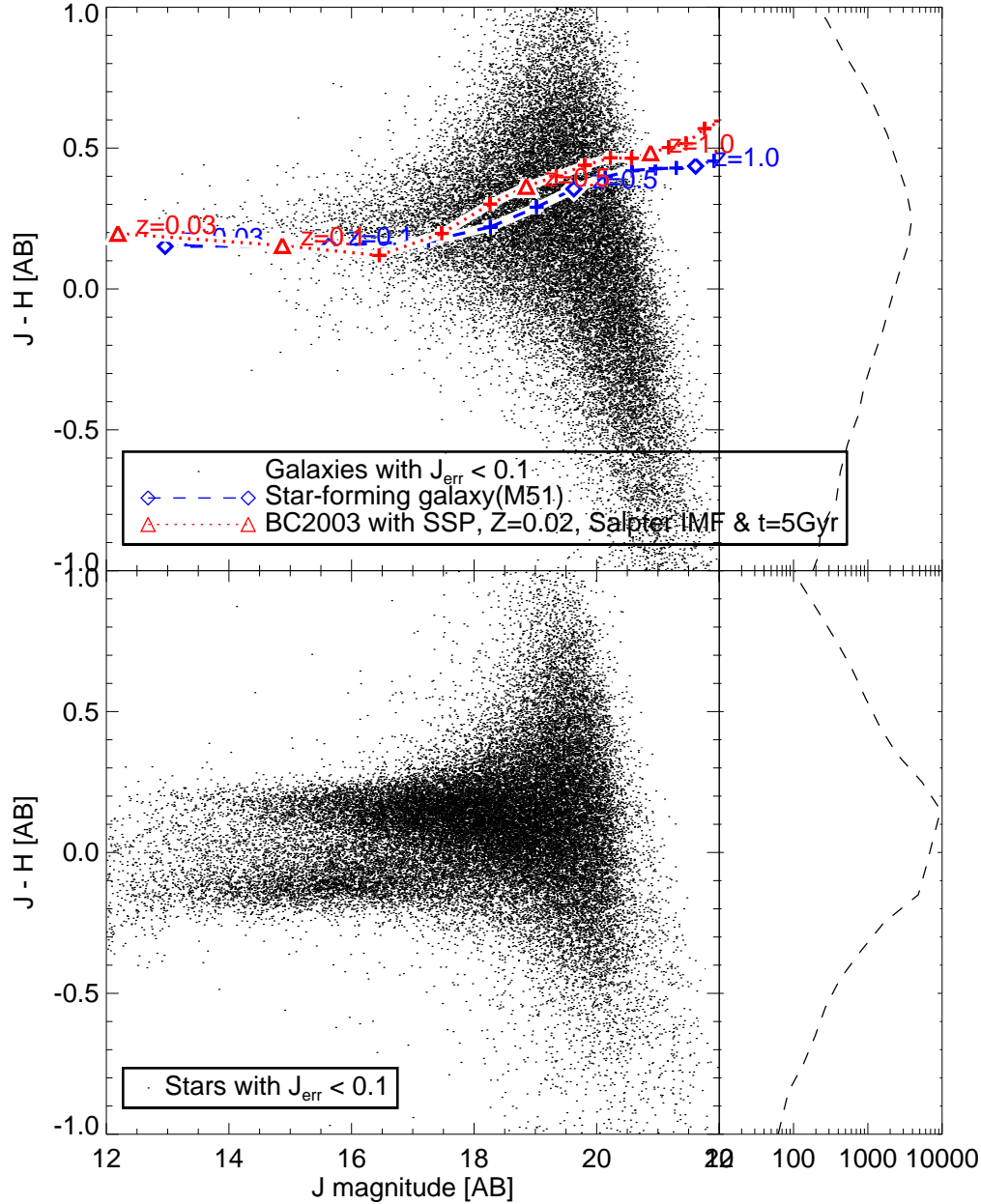}
\caption{ Color-magnitude diagram of $J$ versus $J-H$ (left) and histograms showing the distribution of $J-H$ colors (right). The upper panel shows the galaxies with $J$ magnitude error less than 0.1 mag. We used same condition of galaxy redshift track at Section \ref{sec:galcut}. The redshift track is for galaxies with the characteristic absolute magnitudes. The lower panel shows the distribution and histogram of the stars of the same conditions in the upper panel. 
\label{f12}}
\end{figure*}

The galaxy number count as a function of magnitude is a quantity used in investigating galaxy evolution. Due to the contamination of dust extinction and star formation, the galaxy number counts in NIR bands provide more accurate information of galaxy evolution than those using the optical bands. Here, we show the galaxy number count as a function of magnitude for representation of depth and homogeneity of our survey. The total area used in calculating the galaxy number count is 4.98 deg$^2$, where the optical data \citep{jeon10,hwan07} and the {\it AKARI} IR data overlap. 

Using the galaxies defined so in Section \ref{sec:galcut}, we calculated the number of galaxies per square degree for each magnitude bin. 
Since the depth of each subfield is different from another, we computed the number count for each subfield down to the 99$\%$ reliability magnitudes limit. Then the completeness correction was applied to the number counts for each subfield. Finally, we combined the number counts from all the subfields to derive the final number count result. We also calculated the number count of stars up to the 99$\%$ reliability magnitude. The errors from the number counts are computed from the Poisson errors. 

Figure \ref{f11} shows the galaxy and star number counts in $J$ and $H-$band. 
The lower solid line with crosses represents the raw number counts and the upper solid line with crosses shows the number counts after applying the completeness correction. We compared our number counts with those in the literature \citep{yan98,sara99,sara01,mar01,chen02,iov02,fir06,met06,cri09,kee10}. 
Since the filter systems in the literature do not exactly match with that of our survey and 
also due to the cosmic variance, some offsets may exist.
For example, \citet{met06} mention that their number counts, covering only  0.95$\times$0.95 arcmin$^2$,
can be uncertain by up to a factor of 2 due to the cosmic variance.
We can see that our number counts are consistent with other results. 
The solid line with asterisks shows the number counts of stars. The galaxy and star number count data are provided in Table \ref{t05}.

\subsection{ Color-Magnitude Diagram} \label{sec:cmd}

Figure \ref{f12} shows the color-magnitude diagram of $J$ versus $J-H$ and its histogram. The upper panel shows the distribution for galaxies with $J-$band magnitude error less than 0.1 mag, and the lower panel shows that for stars with the same condition as galaxies in the upper panel. We used the redshift track to check the position of galaxies on the color-magnitude diagram. We used the same conditions for the galaxy redshift track as used in Section \ref{sec:galcut}. These redshift tracks represent galaxies with characteristic absolute magnitude $M^*$. We adopt $M^*-5 \log h$ for $J-$band as $-22.85\pm0.04$ from \citet{jones06}. We calculated the $k$-correction values in a standard manner \citep{hogg02}. The $J-H$ colors of the star forming galaxies and passive evolving galaxies are similar to each other up to redshift $\sim$1. Therefore the galaxies on Figure \ref{f12} are those within the redshift range 0 to 1, and those at higher redshift with magnitude brighter than $M^*$. 

\section{SUMMARY}
We have presented the properties of sources using $J$ and $H-$band data from FLAMINGOS on the KPNO 2.1m telescope, a set of follow-up imaging observations of the {\it AKARI} NEP field. We cover 5.1 deg$^2$ with 5$\sigma$ depth of $\sim$21.6 mag and $\sim$21.3 mag (AB) for $J$ and $H-$band, respectively. We detected 208,020 sources for $J-$band and 203,830 sources for $H-$band. Our NIR data provides multi-wavelength coverage for the {\it AKARI} NEP sources along with the optical and the MIR data, enabling detailed SED analysis. Through these SEDs, we can obtain key physical properties such as stellar mass and photometric redshifts, and select interesting objects such as high redshift galaxies.

\acknowledgments
This work was supported by the National Research Foundation of Korea (NRF) grant, No. 2008-0060544, funded by the Korea government (MSIP). HML acknowledges the support from the grant No. 2012R1A4A1028713 of the National Research Foundation of Korea (NRF) grant funded by the Korea Government (MSIP). HM is supported by a JSPS grant 23244040.

\begin{onecolumngrid}
\clearpage
\begin{landscape}

\begin{deluxetable*}{cccccc}
\tablecolumns{6}
\tablewidth{0pt}
\tablecaption{ Observation summary.\label{t02}}
\tablehead{
\colhead{Filter} & \colhead{Total exposure time [minutes]}  & \colhead{Exposure time per image [minutes]} &
\colhead{Seeing [$\arcsec$]} & \colhead{5$\sigma$ detection limits [AB mag]} 
& \colhead{50$\%$ Completeness [AB mag]} 
}
\startdata
$J$ & 28.5 -- 52.5 (36) & 1.5 & 1.4 -- 2.4 (1.8) & 20.88 -- 21.94 (21.63) & 20.11 -- 21.01 (20.65) \\
$H$ & 11.3 -- 34.0 (21.3) & 0.5 -- 1.0 (0.7) & 1.4 -- 2.6 (1.7) & 20.42 -- 21.72 (21.29) & 19.44 -- 20.81 (20.35)\\
\enddata
\tablecomments{The values in the parentheses are the median values.}
\end{deluxetable*}
\clearpage

\LongTables 
\begin{deluxetable*}{cccccccccccc}
\tablecolumns{12}
\tablewidth{0pc}
\tablecaption{ 
Photometric zeropoint, seeing, 5$\sigma$ detection limit, magnitude of 50$\%$ completeness, and magnitude of 99$\%$ reliability for each subfield.\label{t03}}
\tablehead{
\colhead{} &  \multicolumn{5}{c}{$J-$band} &\colhead{}&  \multicolumn{5}{c}{$H-$band} \\
\cline{2-6} \cline{8-12}  \\
\colhead{Field} & 
\colhead{$Z_p$\tablenotemark{a}}& \colhead{Seeing} & \colhead{detection limits \tablenotemark{b}} & \colhead{comp50$\%$\tablenotemark{c}} & \colhead{relia99$\%$\tablenotemark{d}} & \colhead{} &
\colhead{$Z_p$\tablenotemark{a}}& \colhead{Seeing} & \colhead{detection limits \tablenotemark{b}} & \colhead{comp50$\%$\tablenotemark{c}} & \colhead{relia99$\%$\tablenotemark{d}}  \\
\colhead{} & \colhead{(mag)}& \colhead{($\arcsec$)} & \colhead{(mag)} & \colhead{(mag)} & \colhead{(mag)} & \colhead{} &
\colhead{(mag)}& \colhead{($\arcsec$)} & \colhead{(mag)} & \colhead{(mag)} & \colhead{(mag)} 
} 
\startdata

FLM01&  22.43&   1.67&  21.31&  20.40&  21.46&&  23.22&   1.98&  21.06&  20.06&  21.12 \\
FLM02&  22.41&   1.67&  21.87&  20.85&  21.78&&  23.26&   2.43&  20.93&  19.85&  21.15 \\
FLM03&  22.43&   1.67&  21.85&  20.97&  21.74&&  23.26&   2.58&  20.74&  19.70&  21.11 \\
FLM04&  22.42&   1.67&  21.70&  20.69&  21.63&&  23.22&   1.67&  21.17&  20.18&  21.34 \\
FLM05&  22.45&   1.98&  21.24&  20.20&  21.30&&  23.21&   1.67&  21.08&  20.14&  21.12 \\
FLM06&  22.43&   1.98&  21.32&  20.25&  21.47&&  23.14&   2.28&  20.49&  19.42&  21.01 \\
FLM07&  22.36&   1.98&  21.49&  20.47&  21.48&&  23.16&   1.98&  20.42&  19.58&  20.63 \\
FLM08&  22.46&   1.98&  21.48&  20.49&  21.48&&  23.13&   1.67&  20.96&  19.94&  20.94 \\
FLM09&  22.41&   1.67&  21.74&  20.78&  21.58&&  23.34&   1.82&  20.92&  19.97&  20.94 \\
FLM10&  22.43&   1.67&  21.73&  20.80&  21.68&&  23.21&   1.98&  20.94&  20.08&  21.11 \\
FLM11&  22.42&   1.98&  21.17&  20.09&  21.33&&  23.23&   1.52&  21.28&  20.35&  21.20 \\
FLM12&  22.57&   1.67&  21.72&  20.79&  21.54&&  23.28&   1.52&  21.24&  20.24&  21.14 \\
FLM13&  22.41&   1.82&  21.71&  20.69&  21.69&&  23.22&   1.52&  21.30&  20.37&  21.24 \\
FLM14&  22.71&   1.82&  21.61&  20.70&  21.65&&  23.22&   1.52&  21.31&  20.40&  21.29 \\
FLM15&  22.40&   1.82&  21.76&  20.70&  21.75&&  23.21&   1.67&  21.29&  20.29&  21.16 \\
FLM16&  22.58&   1.98&  21.53&  20.38&  21.64&&  23.27&   1.67&  21.00&  20.02&  21.06 \\
FLM17&  22.48&   1.82&  21.58&  20.66&  21.61&&  23.19&   1.82&  21.01&  19.98&  21.09 \\
FLM18&  22.49&   1.98&  21.34&  20.31&  21.43&&  23.26&   1.98&  20.91&  19.97&  21.08 \\
FLM19&  22.44&   2.13&  21.06&  20.10&  21.25&&  23.25&   1.67&  21.28&  20.33&  21.17 \\
FLM20&  21.98&   1.82&  21.47&  20.36&  21.45&&  23.21&   2.13&  21.17&  20.10&  21.33 \\
FLM21&  22.47&   1.52&  21.63&  20.73&  21.58&&  23.22&   1.98&  21.28&  20.20&  21.14 \\
FLM22&  22.51&   1.52&  21.63&  20.62&  21.51&&  23.27&   1.52&  21.47&  20.46&  21.28 \\
FLM23&  22.47&   1.82&  21.93&  20.98&  21.69&&  23.23&   1.52&  21.48&  20.58&  21.21 \\
FLM24&  22.52&   2.13&  21.68&  20.59&  21.76&&  23.24&   1.67&  21.36&  20.31&  21.27 \\
FLM26&  22.59&   2.43&  21.51&  20.26&  21.65&&  22.81&   1.52&  21.70&  20.71&  21.38 \\
FLM27&  22.53&   1.67&  21.94&  20.93&  21.67&&  23.20&   1.82&  21.68&  20.71&  21.45 \\
FLM28&  22.54&   1.98&  21.25&  20.22&  21.31&&  23.46&   1.52&  21.30&  20.48&  21.15 \\
FLM29&  22.35&   1.67&  21.69&  20.70&  21.70&&  23.36&   1.82&  21.16&  20.19&  21.15 \\
FLM30&  21.78&   1.67&  21.53&  20.34&  21.56&&  23.29&   2.13&  21.12&  20.17&  21.35 \\
FLM31&  22.39&   1.82&  21.50&  20.58&  21.59&&  23.37&   1.52&  21.38&  20.55&  21.20 \\
FLM32&  22.40&   1.67&  21.75&  20.73&  21.59&&  23.23&   1.67&  21.29&  20.42&  21.21 \\
FLM33&  22.43&   1.98&  21.72&  20.78&  21.64&&  23.25&   1.67&  21.45&  20.58&  21.36 \\
FLM34&  22.43&   1.67&  21.73&  20.70&  21.50&&  22.81&   1.67&  21.20&  20.24&  21.14 \\
FLM35&  22.30&   2.13&  21.80&  20.50&  21.78&&  22.84&   1.52&  21.67&  20.64&  21.39 \\
FLM36&  22.61&   2.13&  21.60&  20.53&  21.63&&  23.08&   1.67&  21.23&  20.27&  21.13 \\
FLM37&  22.50&   1.52&  21.62&  20.67&  21.50&&  23.29&   1.67&  21.22&  20.25&  21.21 \\
FLM38&  22.40&   1.52&  21.87&  20.80&  21.44&&  23.29&   1.67&  21.25&  20.18&  21.12 \\
FLM39&  22.39&   1.67&  21.76&  20.72&  21.65&&  23.22&   1.98&  21.19&  20.17&  21.27 \\
FLM40&  22.36&   1.98&  21.41&  20.50&  21.56&&  23.28&   1.67&  21.29&  20.34&  21.20 \\
FLM41&  22.23&   1.82&  21.21&  20.14&  21.29&&  23.23&   1.52&  21.33&  20.48&  21.21 \\
FLM42&  22.44&   1.98&  21.65&  20.66&  21.67&&  23.24&   1.67&  21.32&  20.46&  21.20 \\
FLM43&  22.54&   1.98&  21.65&  20.61&  21.69&&  22.82&   1.67&  21.28&  20.40&  21.22 \\
FLM44&  22.53&   1.82&  21.76&  20.71&  21.79&&  22.81&   1.82&  21.14&  20.22&  21.22 \\
FLM45&  22.40&   1.52&  21.69&  20.77&  21.61&&  22.80&   1.67&  21.48&  20.51&  21.36 \\
FLM46&  22.58&   2.28&  21.50&  20.22&  21.63&&  22.87&   1.67&  21.41&  20.42&  21.22 \\
FLM47&  22.38&   1.82&  21.38&  20.41&  21.45&&  23.27&   1.67&  21.33&  20.34&  21.23 \\
FLM48&  22.41&   1.82&  21.41&  20.37&  21.52&&  23.21&   1.67&  21.35&  20.34&  21.38 \\
FLM49&  22.41&   1.67&  21.55&  20.47&  21.44&&  23.27&   1.67&  21.37&  20.32&  21.34 \\
FLM50&  22.59&   1.67&  21.47&  20.51&  21.50&&  23.20&   1.67&  21.42&  20.42&  21.40 \\
FLM51&  22.42&   1.67&  21.66&  20.76&  21.65&&  23.19&   1.82&  21.38&  20.40&  21.40 \\
FLM52&  22.41&   1.67&  21.72&  20.60&  21.59&&  23.32&   1.67&  21.34&  20.23&  21.25 \\
FLM53&  22.56&   1.67&  21.70&  20.86&  21.68&&  23.17&   1.37&  21.64&  20.80&  21.27 \\
FLM54&  22.56&   1.67&  21.65&  20.62&  21.57&&  23.24&   1.52&  21.42&  20.54&  21.22 \\
FLM55&  22.46&   1.67&  21.80&  20.75&  21.66&&  23.28&   1.67&  21.37&  20.40&  21.30 \\
FLM56&  22.51&   1.82&  21.68&  20.45&  21.79&&  23.19&   1.67&  21.28&  20.25&  21.40 \\
FLM57&  22.49&   1.98&  21.50&  20.44&  21.52&&  23.19&   1.67&  21.36&  20.47&  21.22 \\
FLM58&  22.50&   1.67&  21.51&  20.43&  21.50&&  23.20&   1.37&  21.72&  20.75&  21.38 \\
FLM59&  22.48&   1.67&  21.59&  20.63&  21.53&&  23.20&   1.52&  21.57&  20.61&  21.40 \\
FLM60&  22.49&   1.67&  21.67&  20.68&  21.65&&  23.20&   1.52&  21.59&  20.59&  21.36 \\
FLM61&  22.56&   1.67&  21.75&  20.68&  21.57&&  23.86&   1.52&  21.03&  20.19&  20.96 \\
FLM62&  22.62&   1.67&  21.81&  20.74&  21.71&&  23.32&   1.52&  21.48&  20.60&  21.21 \\
FLM63&  22.52&   1.67&  21.81&  20.78&  21.48&&  23.18&   1.52&  21.50&  20.64&  21.26 \\
FLM64& \nodata& \nodata& \nodata& \nodata& \nodata&&  23.16&   1.82&  21.25&  20.31&  21.20 \\
FLM65& \nodata& \nodata& \nodata& \nodata& \nodata&&  22.79&   1.82&  21.10&  20.12&  21.10 \\
FLM66& \nodata& \nodata& \nodata& \nodata& \nodata&&  22.83&   1.67&  21.32&  20.49&  21.23 \\

\enddata
\tablenotetext{a}{The apparent magnitude is $M=Z_p - 2.5\log$ $(DN/sec)$.}
\tablenotetext{b}{The 5$\sigma$ detection limit for a circular aperture with diameter of 3 times FWHM}
\tablenotetext{c}{The magnitude corresponding to the 50$\%$ completeness}
\tablenotetext{d}{The magnitude corresponding to the 99$\%$ reliability}
\end{deluxetable*}
\clearpage

\begin{deluxetable*}{ccccccccccccccc}
\setlength{\tabcolsep}{0.02in}
\tablecaption{$J$ and $H-$band merged catalog.\label{t04}}
\tablewidth{0pt}
\tablehead{
\colhead{ID} & \colhead{R.A.} & \colhead{Decl.} & \colhead{$J_{AUTO}$} & \colhead{Error} & 
\colhead{$J_{APER}$} & \colhead{Error} & \colhead{$H_{AUTO}$} & \colhead{Error} & 
\colhead{$H_{APER}$} & \colhead{Error}  
& \colhead{Stellarity} & \colhead{$A_J$} & 
\colhead{$A_H$} & \colhead{flag}\\
\colhead{(1)} & \colhead{(2)} & \colhead{(3)} & \colhead{(4)} & \colhead{(5)} & 
\colhead{(6)} & \colhead{(7)} & \colhead{(8)} & \colhead{(9)} & 
\colhead{(10)} & \colhead{(11)}  
& \colhead{(12)} & \colhead{(13)} & 
\colhead{(14)} & \colhead{(15)}
}
\startdata

FLM01\_00651&18:03:16.74&+67:32:13.20& 20.911&  0.136& 20.966&  0.168& 20.885&  0.177& 21.299&  0.276& 0.65& 0.045& 0.029&\\
FLM01\_00652&18:03:16.75&+67:41:18.77&    $NaN$&    $NaN$&    $NaN$&    $NaN$& 20.838&  0.202& 20.746&  0.172& 0.47&   $NaN$& 0.027&\\
FLM01\_00653&18:03:16.81&+67:22:15.77& 21.172&  0.199& 21.199&  0.200& 99.999& 99.999& 99.999& 99.999& 0.48& 0.045&   $NaN$&\\
FLM01\_00654&18:03:16.88&+67:37:10.82& 20.198&  0.074& 20.109&  0.092& 99.999& 99.999& 99.999& 99.999& 0.67& 0.042&   $NaN$&\\
FLM01\_00655&18:03:16.91&+67:36:12.73& 99.999& 99.999& 99.999& 99.999& 20.715&  0.158& 20.921&  0.199& 0.49&   $NaN$& 0.027&\\
FLM01\_00656&18:03:16.92&+67:30:57.40& 20.254&  0.086& 20.342&  0.106& 20.400&  0.112& 20.464&  0.137& 0.89& 0.046& 0.029&\\
FLM01\_00657&18:03:16.96&+67:33:30.66& 20.399&  0.103& 20.524&  0.120& 19.795&  0.100& 19.798&  0.087& 0.16& 0.044& 0.028&\\
FLM01\_00658&18:03:16.97&+67:26:35.22& 20.136&  0.072& 20.265&  0.101& 99.999& 99.999& 99.999& 99.999& 0.90& 0.046&   $NaN$&\\
FLM01\_00659&18:03:17.05&+67:27:24.29& 99.999& 99.999& 99.999& 99.999& 21.049&  0.177& 21.198&  0.253& 0.61&   $NaN$& 0.030&\\
FLM01\_00660&18:03:17.05&+67:37:27.81& 20.610&  0.184& 20.866&  0.155& 99.999& 99.999& 99.999& 99.999& 0.55& 0.042&   $NaN$&\\
FLM01\_00661&18:03:17.09&+67:36:35.88& 21.122&  0.187& 21.149&  0.195& 99.999& 99.999& 99.999& 99.999& 0.58& 0.042&   $NaN$&\\
FLM01\_00662&18:03:17.10&+67:37:22.53& 21.117&  0.217& 21.255&  0.214& 99.999& 99.999& 99.999& 99.999& 0.65& 0.042&   $NaN$&\\
FLM01\_00663&18:03:17.15&+67:38:35.97& 20.786&  0.155& 20.829&  0.151& 99.999& 99.999& 99.999& 99.999& 0.51& 0.041&   $NaN$&\\
FLM01\_00664&18:03:17.17&+67:29:08.37& 19.167&  0.070& 19.209&  0.066& 19.016&  0.067& 19.050&  0.065& 0.89& 0.046& 0.030&\\
FLM01\_00665&18:03:17.19&+67:41:50.41&    $NaN$&    $NaN$&    $NaN$&    $NaN$& 18.020&  0.058& 18.138&  0.057& 0.17&   $NaN$& 0.027&\\
FLM01\_00666&18:03:17.22&+67:34:37.53& 20.523&  0.148& 20.908&  0.160& 99.999& 99.999& 99.999& 99.999& 0.34& 0.043&   $NaN$&\\
FLM01\_00667&18:03:17.23&+67:32:17.26& 20.957&  0.135& 21.282&  0.219& 20.929&  0.197& 20.760&  0.174& 0.40& 0.045& 0.029&\\
FLM01\_00668&18:03:17.23&+67:33:04.26& 20.589&  0.161& 20.984&  0.170& 99.999& 99.999& 99.999& 99.999& 0.47& 0.044&   $NaN$&\\
FLM01\_00669&18:03:17.24&+67:37:29.69& 99.999& 99.999& 99.999& 99.999& 19.754&  0.106& 20.266&  0.118& 0.49&   $NaN$& 0.026&\\
FLM01\_00670&18:03:17.26&+67:39:22.08& 21.222&  0.188& 21.120&  0.191& 99.999& 99.999& 99.999& 99.999& 0.65& 0.042&   $NaN$&\\
FLM01\_00671&18:03:17.31&+67:45:01.33&    $NaN$&    $NaN$&    $NaN$&    $NaN$& 20.456&  0.114& 20.455&  0.136& 0.37&   $NaN$& 0.028&\\
FLM01\_00672&18:03:17.35&+67:40:17.33& 21.132&  0.146& 21.319&  0.226& 99.999& 99.999& 99.999& 99.999& 0.49& 0.042&   $NaN$&\\
FLM01\_00673&18:03:17.42&+67:35:23.47& 21.448&  0.142& 23.891&  2.331& 99.999& 99.999& 99.999& 99.999& 0.52& 0.042&   $NaN$&\\
FLM01\_00674&18:03:17.52&+67:23:31.96& 20.646&  0.077& 20.710&  0.137& 99.999& 99.999& 99.999& 99.999& 0.70& 0.045&   $NaN$&\\
FLM01\_00675&18:03:17.58&+67:34:33.18& 20.243&  0.142& 20.824&  0.150& 99.999& 99.999& 99.999& 99.999& 0.65& 0.043&   $NaN$&\\
FLM01\_00676&18:03:17.60&+67:35:51.45& 19.728&  0.079& 20.022&  0.088& 99.999& 99.999& 99.999& 99.999& 0.97& 0.042&   $NaN$&$crt$\\
FLM01\_00677&18:03:17.61&+67:30:59.62& 99.999& 99.999& 99.999& 99.999& 21.111&  0.144& 22.253&  0.654& 0.49&   $NaN$& 0.029&\\
FLM01\_00678&18:03:17.62&+67:33:00.44& 99.999& 99.999& 99.999& 99.999& 21.034&  0.162& 21.097&  0.231& 0.54&   $NaN$& 0.028&\\
FLM01\_00679&18:03:17.63&+67:30:23.69& 21.848&  0.234& 22.219&  0.503& 20.091&  0.111& 20.073&  0.103& 0.52& 0.046& 0.029&\\
FLM01\_00680&18:03:17.67&+67:39:53.53& 17.777&  0.061& 18.106&  0.059& 17.814&  0.058& 17.948&  0.056& 0.30& 0.042& 0.027&\\
\enddata
\tablecomments{
(1) The identification number of the source (ID). 
Each ID consists of the subfield name (Figure \ref{f02}) and number in the order of their R.A..\\
(2) and (3) The J2000.0 R.A. and Decl. of a source in a sexigesimal.\\
(4)--(11) are the total magnitude, its magnitude error, aperture magnitude and its magnitude error of a source 
in the $J$ and $H-$band. The aperture magnitudes are obtained with the aperture diameter of 3 times FWHM.
The magnitudes are not corrected for the galactic extinctions.
We used a dummy value of 99.999 for non-detections. If the object has only single filter data, we used {\it NaN} for the other filter information.\\ 
(12) The stellarity index of a source from SE. See Figure \ref{f08} for the distribution of this index. \\
(13--14) The galactic extinction values of $J$ and $H-$band from extinction map of \citet{schl98}. \\
(15) The flag parameter. It contains the crosstalk information ({\it crt}). See section \ref{sec:crstk} for details .
}
\end{deluxetable*}
\clearpage

\end{landscape}
\end{onecolumngrid}

\end{document}